\documentstyle[epsf,notoc]{JHEP}
\def\ifb{{\rm fb}^{-1}}
\def\sbma{\sin(\beta-\alpha)}
\def\s2bma{\sin^2(\beta-\alpha)}
\def\cbma{\cos(\beta-\alpha)}
\def\c2bma{\cos^2(\beta-\alpha)}
\def\mh{m_h}
\def\mH{m_H}
\def\tgg{\to\gamma\gamma}
\def\tbb{\to b\bar b}

\def\simlt{\stackrel{<}{{}_\sim}}
\def\simgt{\stackrel{>}{{}_\sim}}
\def\bea{\begin{eqnarray}}
\def\eea{\end{eqnarray}}
\def\bce{\begin{centering}}
\def\ece{\end{centering}}
\def\bit{\begin{itemize}}
\def\eit{\end{itemize}}
\preprint{ANL--HEP--PR--99--79 \\ CERN--TH/99--203} 

\title{The complementarity of LEP, the Tevatron and the LHC in the search for 
a light MSSM Higgs boson}
\author{M. Carena \\ Theory Division, CERN,
1211 Geneva 23, Switzerland 
\thanks{On leave of absence from
Fermi National Accelerator Laboratory, Batavia, IL 60510, USA} \\
\email{carena@fnal.gov}}
\author{S. Mrenna \\
Physics Department, University of California at Davis,
Davis, CA  95616, USA \\
\email{mrenna@physics.ucdavis.edu}}
\author{C.E.M. Wagner \\
High Energy Physics Division, Argonne National Laboratory, Argonne, 
IL 60439, USA 
\thanks{On leave of absence from
CERN, 1211 Geneva 23, Switzerland} \\
\email{Carlos.Wagner@cern.ch}}

\abstract{
We study the properties of the Higgs boson sector in the MSSM,
putting special emphasis on radiative effects which can affect
the discovery potential of the LHC, Tevatron and/or LEP colliders.
We concentrate on the $Vb\bar{b}$ channel, with $V= Z$ or $W$, and
on the channels with diphoton final states, which are the dominant
ones for the search for a light Standard Model Higgs boson at
LEP/Tevatron and LHC, respectively.
By analyzing the regions of parameter space for
which the searches in at least one of these colliders can be particularly
difficult, we demonstrate the complementarity of these three colliders
in the search for a light Higgs boson which couples in a relevant way
to the $W$ and $Z$ gauge bosons (and hence plays a relevant role in
the mechanism of electroweak symmetry breaking).}

\begin{document}

\section{Introduction}
\label{sec:intro}

The Standard Model (SM) of particle physics provides an excellent 
description of data from collider experiments, including the 
precision electroweak
observables measured at LEP and SLD. The fit to the data 
clearly improves 
if the Higgs boson has a mass less than 250 GeV~\cite{ewfit}. 
LEP is the only accelerator currently running which can directly
test for the existence of a Standard Model-like Higgs boson, if
its mass is sufficiently light~\cite{CARZER}.
The LEP experiments at CERN have recently performed searches for a
Standard Model Higgs boson at a center of 
mass energy of $\sqrt{s} = 189$ GeV. Preliminary limits on the 
Higgs mass of about 95 GeV were set
by two of the experiments, {\sc DELPHI} and {\sc L3}. The
other two experiments, {\sc ALEPH} and {\sc OPAL}, see a small 
excess of events in the
mass window 90--96 GeV~\cite{Felcini} and set an exclusion
limit of about 91 GeV. 
LEP is currently running at a center of mass energy
of 192--196 GeV and will increase gradually the center--of--mass energy
towards 200 GeV and collect data until the end of the year 2000.

In spite of the phenomenological success of the SM, an explanation
of the hierarchy between the Planck and the electroweak scales can
only be obtained if new physics is present at scales of the order of
the weak scale. The success of the SM in describing the precision 
electroweak data suggests (although it does not require)
that any new physics should be weakly--coupled, 
should lead to small or negligible
corrections to precision electroweak observables,
and, in addition, should be consistent with a light Higgs boson.
Low energy supersymmetry provides such an extension of the Standard
Model.

In the minimal supersymmetric extension of the Standard Model (MSSM),
the Higgs sector contains two doublets. At tree-level, the down
and up quarks only couple to the neutral components of the Higgs
doublet $H_1$ and $H_2$, respectively, preventing dangerous 
flavor--changing neutral current (FCNC)
effects.   The ratio of the two Higgs
doublet expectation values, $v_1$ and $v_2$,  
is parametrized by $\tan\beta = v_2/v_1$. 
The Higgs spectrum
consists of one charged, $H^{\pm}$,  
one CP-odd, $A$, and
two CP-even, $h$ and $H$, Higgs bosons.  
At tree-level all Higgs boson masses may be
expressed as a function of  $\tan\beta$, $m_A$ and the $W$ and $Z$ boson
masses, and an upper bound on the lightest CP-even Higgs mass
is found, $m_h \leq M_Z |\cos 2 \beta|$. This bound is modified by
radiative corrections, which depend quartically on the top quark mass
and logarithmically on the stop 
masses~\cite{HEHO,CAESQURI,CAESQUWA,HHH,HEHOWE}.  
As will be discussed below, even after the inclusion of radiative
corrections, an upper bound on the lightest CP-even
Higgs mass is obtained for large
values of the CP-odd Higgs mass $m_A \simgt 300$ GeV, 
for which the lightest CP-even Higgs boson has standard
model-like properties. 

The supersymmetric spectrum is constrained by
direct experimental searches and by the requirement that it provides a
good description of the precision electroweak data. This requirement
implies that, unless unnatural cancellations take 
place~\cite{HERA},
the soft supersymmetry-breaking mass parameter
for the left-handed top-squark  should 
be larger than  300 GeV. Quite generally,
the heavier the supersymmetric spectrum, 
and in particular the heavier the left-handed sfermions, 
the better the agreement between the MSSM and
the precision electroweak observables. 
If supersymmetric particles are heavy, the 
low energy properties  of the Higgs
sector of the MSSM can be described by an effective theory
containing two Higgs
doublets, with couplings and masses fixed
by the proper  matching conditions at the 
scale of the supersymmetric particle masses.  

In this low energy, effective theory, the couplings of the two CP-even
Higgs bosons, $h$ and $H$,  to the $W$ and $Z$ bosons are given by the 
SM Higgs couplings multiplied by
$\sin (\alpha - \beta)$ and $\cos (\alpha - \beta)$, respectively,
where $\alpha$ is the Higgs mixing angle. These scaling factors 
are just the projections of
the CP-even Higgs bosons on $\Phi$, defined as
the Higgs combination that acquires
vacuum expectation value, 
\begin{equation}
\Phi = \sqrt{2} \left[Re(H_1^0) \cos\beta + Re(H_2^0) \sin \beta \right]
\equiv v \; 
+h\sbma+H\cbma
\end{equation}
where $v\simeq 246$ GeV is the SM Higgs vacuum expectation value.
In the MSSM, $\Phi$ is not a mass eigenstate. 
However, $\sin(\alpha - \beta)$ or $\cos(\alpha - \beta)$ 
becomes close to one in large
regions of parameter space, reflecting the fact that one of the
two Higgs bosons is mainly responsible for the electroweak symmetry
breaking. In particular, for relatively large values of 
the CP-odd Higgs mass ($m_A \geq 300$ GeV), one finds
$\sin (\alpha - \beta) \approx 1$. 

The Higgs searches at LEP, the Tevatron and the Large Hadron Collider (LHC)
are motivated by the desire to understand the mechanism of 
electroweak symmetry
breaking. For this reason, 
it is most important to find the Higgs with
relevant couplings to the $W$ and $Z$ bosons, with $\s2bma$
(or $\c2bma$) close to unity.  For convenience, we denote such 
a Higgs boson as $\phi_W$ to emphasize its couplings to the $W$ (and $Z$).
As shown in Appendix A, there is
a useful relation between the masses of the CP-even Higgs bosons and
their couplings to the $W$ and $Z$ bosons,
\begin{equation}
\left.
m_h^2 \; \sin^2(\beta - \alpha) + m_H^2 \; \cos^2 (\beta - \alpha) 
= m_h^2 
\right|_{m_A \gg M_Z}.   
\label{relation}
\end{equation}
In Eq.~(\ref{relation}),
the right hand side is equal to the upper
bound on the Higgs boson mass, which, for
squark masses of the order of 1 TeV,
is about 120--130 GeV for moderate or
large values of $\tan\beta$ and about 100 GeV
for $\tan\beta$ close to one~\cite{CAESQUWA,CACHPOWA}. 
The above relation, Eq.~(\ref{relation}), implies
that the Higgs $\phi_W$ that couples in a relevant way to the 
$W$ and $Z$ bosons
should also be relatively light, with a mass
close to the upper bound when it couples to the $W$ and $Z$ boson
with nearly SM strength.
Therefore, not only is it true that the lightest CP--even Higgs boson
mass is bounded from above, but, when $\cos(\beta-\alpha) \approx 1$, 
then the bound applies
to $\mH$, with $\mh$ being even smaller.
If $\s2bma\to 1$ or $\c2bma\to 1$, there can be a substantial
mass splitting between $h$ and $H$.  However, in the latter case,
the mass splitting cannot be too large, or
the lightest CP-even Higgs boson would have been already seen at LEP.   

\section{Higgs searches at present and (near) future colliders}
\label{sec:searches}

There are three experiments that are expected to
search for the Higgs $\phi_W$
in the mass range 95--130 GeV. LEP, at present, 
the Tevatron in the years 2000--2006 and 
LHC from 2005 on. As we mentioned 
above, LEP is actively looking for a Higgs boson with 
couplings to the $Z$ boson and with a mass below or near
100 GeV in the channel $Z \phi_W$ with  
$\phi_W \rightarrow b \bar{b}$  or $\tau^+ \tau^-$.
If LEP reaches
a center--of--mass energy of about 200 GeV and collects 
200 pb$^{-1}$ of data per experiment, then
evidence for such a Higgs boson should be observed if:
\begin{enumerate}
\item $\sin^2(\alpha-\beta)\simeq {\cal O}(1)$ 
(or $\cos^2(\alpha-\beta)\simeq {\cal O}(1)$), \\
\item The CP-even Higgs mass 
$m_h \simlt 107$ GeV (or $m_H \simlt 107$ GeV), \\
\item The branching ratio
BR($h\rightarrow b\bar{b}$)  (or BR$(H\rightarrow b\bar{b})$) is
close to the Standard Model value.
\end{enumerate}

The Run II of the 
Tevatron Collider is expected to
start in the year 2000. The
Tevatron will have sensitivity to Higgs boson in the $V\phi$ channel, with
$V = Z$ or $W$, and $\phi \rightarrow b \bar{b}$. Hence, its
discovery potential depends also on points number 1 and 3 above,
although the kinematic constraint on the Higgs mass may be relaxed
(point 2).
However, the Tevatron discovery potential will depend strongly
on the final integrated luminosity collected by the CDF and
D0 experiments~\cite{SHWG}.  

Experiments at the LHC will rely mainly on the 
signature
$pp\to\gamma\gamma+X$ to detect a Higgs $\phi_W$
in the mass range $m_{\phi_W}\simlt 130$ GeV. In particular,
the {\sc ATLAS} and {\sc CMS} experiments have performed studies that show
sensitivity to $\phi_W$ in the channels $gg\to\phi(\tgg)$,
$t\bar t\phi(\tgg)$, $W\phi(\tgg)$, and $t\bar t\phi(\tbb)$.\footnote{
Several other channels have been proposed which would be useful for
studying a Higgs boson
with SM-like couplings to the gauge bosons and
up-quarks\cite{wwfusion,stophiggs}.  However, these
are experimentally challenging, and, to the best of
our knowledge, the {\sc ATLAS} and {\sc CMS} collaborations have not
yet analyzed the reach in these channels. We shall therefore not discuss 
these channels in detail, although we shall analyze their 
possible relevance in sections~\ref{sec:couplings} 
and~\ref{subsec:bottaucorr}.} 
These studies show that these channels cover wide
regions of the $m_A-\tan\beta$ plane of the MSSM, with the small
$m_A$ region ($m_A \simlt 250$ GeV) being the most difficult.
Larger coverage of
the $m_A-\tan\beta$ plane in the MSSM can be achieved by considering
also the production and decay signatures of all MSSM Higgs
bosons~\cite{RICHTER,CMS}.  In these
analyses, it is assumed that the sparticles have typical masses $M_S$ of
order 1 TeV, and that the stop trilinear coupling 
$\tilde{A}_t = A_t - \mu/\tan\beta$ is much smaller than $M_S$.
The latter is, in principle,
a conservative assumption, since, for low luminosity, 
and for a Higgs mass $m_{\phi_W} \leq 130$ GeV, the reach
potential improves for larger values of the Higgs mass (large
values of $\tilde{A}_t, M_S$, etc.). In the present study, we
are interested in the search for $\phi_W$ and we shall hence
concentrate on only its signatures.

  The object
of the present study is to illustrate the relationship between
measurements at the different colliders, and to demonstrate 
the potential of the combined experimental program to discover
the Higgs $\phi_W$. 
In the following section, we will review the behavior of the
Higgs boson couplings to particles and sparticles with respect
to variations in the MSSM parameters.  In particular, we will
show their impact on production cross sections and 
branching ratios.  Given the sensitivity of the various experiments
to discover a Higgs boson as a function of the Higgs boson mass and
their integrated luminosity,
we then calculate the corresponding sensitivity in the MSSM, based on the
Standard Model experimental simulations done at the LEP,
Tevatron and LHC colliders.\footnote{The
sensitivity or $R$--value is the ratio of the Higgs production
cross section times Higgs decay branching ratio to the ones
necessary to claim discovery in the Standard Model.
If $R>1$, then an enhancement of the Higgs production
rate and/or branching ratio  over the Standard Model
expectation is needed to claim discovery for a stated luminosity.  
If $R<1$, then the Standard Model Higgs boson can
be discovered with less luminosity.}
This will be shown in Section \ref{sec:results}.
We pay particular attention to 
choices of MSSM parameters which will clearly lead to
difficulty at one of the experiments, and explain why this
increases the sensitivity of the complementary experiment.
Our conclusions are stated in Section \ref{sec:conclusions}.

\section{The couplings of the CP-even Higgs bosons}
\label{sec:couplings}

Quite generally, the two CP--even Higgs boson eigenstates are
a mixture of the real, neutral components of the $H_1$ and $H_2$
Higgs doublets, 
\begin{eqnarray}
\left( \begin{array}{c} h \\ H \end{array} \right)= 
\left( \begin{array}{cc} -\sin\alpha & \cos\alpha \\
\cos\alpha & \sin\alpha \end{array} \right)
\left( \begin{array}{c} \sqrt{2} Re(H_1^0) - v_1 \\ 
\sqrt{2} Re(H_2^0) - v_2 \end{array} \right),
\label{mixings}
\end{eqnarray}
and the lightest CP--even Higgs boson couples to down quarks/leptons
and up quarks by its Standard Model values
times  $-\sin\alpha/\cos\beta$ and $\cos\alpha/\sin\beta$, respectively.
The couplings to the heavier CP--even Higgs boson are given by the standard 
model values
times $\cos\alpha/\cos\beta$ and $\sin\alpha/\sin\beta$, respectively.
Analogously, the coupling of the CP--odd Higgs boson to down quarks/leptons
and up quarks is
given by the Standard Model coupling times $\tan\beta$
and $1/\tan\beta$, respectively. The
lightest (heaviest) CP--even Higgs boson has  $VVh$
($VVH$) couplings
which are given by the Standard Model value times $\sin(\beta - \alpha)$
($\cos(\beta - \alpha)$), where $V$ represents a $W$ or $Z$ boson.
The coupling of a CP--even and 
a CP--odd Higgs boson with a $Z$ boson $ZhA$ ($ZHA$) is proportional
to $\cos(\beta - \alpha)$ ($\sin(\beta - \alpha)$). 

As stated above, LEP is presently exploring the Higgs mass region 
95 GeV $\simlt m_{\phi_W} \simlt$ 107 GeV. 
Already the present
bounds on a SM-like Higgs mass, of about 95 GeV, put strong
constraints on the realization of the infrared fixed--point
scenario in the MSSM, and, in general, of the small $\tan\beta$ scenario,
with $\tan\beta$ close to one~\cite{CACHPOWA}. Indeed, if
a SM-like Higgs were discovered at LEP in this region of
mass, the fixed--point scenario could only be 
accommodated for large values
of the stop masses and of the stop mixing parameters. 
In general, this Higgs $\phi_W$ mass range is
naturally obtained for 
$2 < \tan\beta \simlt 5$, with smaller
values of the Higgs mass being obtained for smaller values
of $\tan\beta$. 
Larger values of the ratio of Higgs vacuum expectation
values, $\tan\beta > 5$, tend to lead to values of the
Higgs $\phi_W$ mass beyond the reach of LEP.
It is important to stress that, in the 
presence of large mixing in the lepton sector, as suggested
by the SuperKamiokande data, these
values of $\tan\beta$, $2 < \tan\beta \simlt 5$, are 
consistent with the unification of the bottom and $\tau$ Yukawa 
couplings at the grand unification scale~\cite{CELW}.

As explained above, the present experimental constraints
are most naturally satisfied for moderate or large values
of $\tan\beta$, $\tan\beta > 2$.
For $\tan\beta$ larger than a few, the approximation $\sin\beta \simeq 1$, 
and hence, $1/\cos\beta \simeq \tan\beta$, is a good one, 
and one can construct
the following table of simplified {\it tree level} couplings of fermions 
and gauge bosons ($V\equiv W~{\rm or}~Z$)
to the CP--even Higgs bosons relative to their Standard Model values:
\begin{eqnarray}
\begin{tabular}{||c|c|c|c||} \hline
        & ~~~~$b\bar b$~~~~     & ~~~~$t\bar t$~~~~    
 & ~~~~$VV$~~~~ \\ \hline
~~~~$h$~~~~     & $-\sin\alpha\times \tan\beta$       
& $\cos\alpha$     & $\cos\alpha$ \\
$H$     & $ \cos\alpha\times \tan\beta$       
& $\sin\alpha$     & $\sin\alpha$ \\ \hline
\end{tabular}
\end{eqnarray}
Note that the tree level couplings $t\bar t\phi$ and $VV\phi$ exhibit the
same behavior, so that, for the same values of the Higgs mass the
production rates for $Z\phi$ at LEP, $W/Z\phi$ at the Tevatron,
and $t\bar t/W\phi$ at the LHC
are simultaneously enhanced or suppressed with respect to the Standard
Model case. 
Also, for heavy sparticles, the $\phi \to gg$ 
decay rate (which determines the $gg\to\phi$ production rate)
is approximately proportional to the tree level $t\bar t\phi$ coupling.
Therefore, the production of $t\bar t\phi, W\phi$ and $gg\to\phi$
have the same dependence on the MSSM couplings (when the sparticles
are heavy).  
Moreover, when sparticles are heavy,
the partial width for the decay of the Higgs boson to a photon pair 
depends on one--loop contributions from the top quark and $W$ boson, which
come with opposite signs.
When the sparticles are light, however, all color--charged sparticles affect
the $\phi gg$ coupling, while all electrically--charged sparticles affect
the $\phi\gamma\gamma$ coupling, and we will show a few examples
in which their effect become relevant.
Finally, it is important to remark that,
while for $\tan\beta$ larger than a few,
the $t\bar t\phi$ and $VV\phi$ couplings depend only weakly
on $\tan\beta$, the
$\phi b\bar b$ coupling has a strong dependence on this parameter,
and may be strongly affected by radiative corrections
proportional to $\tan\beta$~\cite{sola,CMW}.
This can have a significant impact on Higgs decay branching ratios.

The LEP experiments are sensitive to the Higgs $\phi_W$
mainly through the $Z\phi(\tbb)$ process.  Given the cross section
limit for the Standard Model Higgs boson, the MSSM limit is
derived by properly including the MSSM couplings.  
Figure 1 shows the reach of the Standard Model Higgs 
discovery potential of the LEP and Tevatron colliders for 
different integrated luminosities of the latter as a function
of $R$, defined as the total $V b \bar{b}$ production rate
normalized to the Standard Model value:
\begin{equation}
R(m_\phi) = \frac{\sigma(V\phi)  \;{\rm BR}(\phi \rightarrow b\bar{b})}
{\sigma(V\phi)_{SM} \; {\rm BR}(\phi \rightarrow b\bar{b})_{SM}}.
\label{r_vvphi}
\end{equation}
The subscript $SM$ in Eq.~(\ref{r_vvphi}) denotes the Standard Model values.
For the case of the MSSM, the production cross section is only
modified by the strength of the $VV\phi$ coupling, 
so that 
\bea
R(m_\phi) = \left\{\s2bma,\c2bma\right\} {{\rm BR}(\phi\tbb)\over 
{\rm BR}(\phi\tbb)_{SM}}
\nonumber\\
\simeq \left\{\cos^2\alpha,\sin^2\alpha\right\} 
{{\rm BR}(\phi\tbb)\over {\rm BR}(\phi\tbb)_{SM}},
\label{rone}
\eea
where the left (right) portion of the expression in brackets refers
to $\phi=h (H)$, and we have used the approximations of Table I in
the second line.  If $R$ is too small, then discovery of a SM--like
Higgs boson will not be possible.  Problem regions are when: $(a)$ 
$h \equiv \phi_W$ or $H \equiv \phi_W$  but 
$\mh$ or $\mH$ is too large to be
kinematically accessible, $(b)$ $h$ ($H$) has SM--like
couplings to the gauge bosons, but ${\rm BR}(\phi_W\tbb)$
is suppressed because the mixing angle $\alpha$ is such that
$\sin\alpha/\cos\beta\ll 1$ ($\cos\alpha/\cos\beta \ll 1$)
or because there are large  
radiative corrections, induced by supersymmetric particles,
which lead to a suppression of the renormalized
$\phi_W b\bar b$ coupling, and $(c)$
$\c2bma\simeq\s2bma$ and $\mh$ and $\mH$ are sufficiently different
in mass, so that the production cross section for both Higgs bosons
is small and the two signals do not overlap.
The experiments in Run II and Run III (the proposed high--luminosity
run) at the Tevatron are sensitive
to both the $W\phi(\tbb)$ and $Z\phi(\tbb)$ processes.  Since these
depend on the same couplings, the previous discussion for LEP holds,
except that $(a)$ will not occur provided the experiments receive
{\it enough integrated luminosity}.

As stated above,
the $\phi b\bar b$ coupling can become small because a tree level coupling
vanishes or because of large radiative corrections.
The former occurs when $\sin\alpha$ or $\cos\alpha$ vanishes, whereas,
as explained below, the latter depends on the soft 
supersymmetry-breaking parameters, but 
tends to occur for small values
of $\sin\alpha$
or $\cos\alpha$, such that the tree level coupling is non-vanishing,
but still suppressed compared to the SM one.
The value of $\alpha$ is determined by diagonalizing the quadratic
mass matrix ${\cal M}^2$ for the CP--even Higgs bosons:
\begin{eqnarray}
{\cal M}^2 = 
\left( \begin{array}{cc} {\cal M}^2_{11} & {\cal M}^2_{12} \\
{\cal M}^2_{12} & {\cal M}^2_{22} \end{array} \right),
\label{matel}
\end{eqnarray}
where the matrix components are given by~\cite{CAESQUWA,CMW}
\begin{eqnarray}
{\cal M}^2_{11} & \simeq & m_A^2 \sin^2\beta + M_Z^2 \cos^2\beta
\nonumber\\
& - &\frac{h_t^4 v^2}{16 \pi^2 } \bar{\mu}^2 \sin^2\beta\tilde{a}^2
\left[ 1 + \frac{t}{16 \pi^2}\left( 6 h_t^2 - 2 h_b^2 - 16 g_3^2
\right) \right]
+ {\cal O}(h_t^2 M_Z^2)
\nonumber\\
& - &\frac{h_b^4 v^2}{16 \pi^2 } \bar{\mu}^2 \sin^2\beta \bar{A}_b^2
\left[ 1 + \frac{t}{16 \pi^2}\left( 6 h_b^2 - 2 h_t^2 - 16 g_3^2
\right) \right]
\nonumber\\
{\cal M}^2_{22} & \simeq & m_A^2 \cos^2\beta + M_Z^2 \sin^2\beta
\left(1 - \frac{3}{8 \pi^2} h_t^2 t \right)
\nonumber\\
& + &\frac{h_t^4 v^2}{16 \pi^2} 12 \sin^2\beta \left\{ 
t \left[ 1 + \frac{t}{16 \pi^2} \left( 1.5 h_t^2 + 0.5 h_b^2 -
8 g_3^2 \right)
\right]
\right.
\nonumber\\
& + & \left. 
\bar{A}_t\tilde{a}\left(1  - {\bar{A}_t\tilde{a} \over 12}\right)
\left[ 1 + \frac{t}{16 \pi^2} \left( 3 h_t^2 +  h_b^2 - 16 g_3^2 \right)
\right] \right\}
\nonumber\\
& - &   \frac{v^2 h_b^4}{16 \pi^2} \sin^2\beta \bar{\mu}^4
\left[ 1 + \frac{t}{16 \pi^2} \left( 9 h_b^2 - 5 h_t^2 - 16 g_3^2 \right)
\right]  
+ {\cal O}(h_t^2 M_Z^2)
\nonumber\\
{\cal M}^2_{12} & \simeq & -\left[m_A^2 + M_Z^2
- \frac{h_t^4 v^2}{8 \pi^2} \left(3 \bar{\mu}^2 - \bar{\mu}^2 \bar{A}_t^2 
\right)\right] 
\sin\beta \cos\beta
\nonumber\\
& + & \left[
\frac{h_t^4 v^2}{16\pi^2}\sin^2\beta \bar\mu \tilde{a}\left[\bar{A}_t
\tilde{a} - 6 \right] + \frac{3 h_t^2 M_Z^2}{32 \pi^2} 
\bar{\mu} \tilde{a} \right] 
\left[ 1 + \frac{t}{16 \pi^2} \left( 4.5 h_t^2 - 0.5 h_b^2 - 16 g_3^2 \right)
\right]
\nonumber\\
& + & 
\frac{h_b^4 v^2}{16\pi^2}\sin^2\beta \bar\mu^3 \bar{A}_b 
\left[ 1 + \frac{t}{16 \pi^2} \left( 7.5 h_b^2 - 3.5 h_t^2 - 16 g_3^2 \right)
\right].
\label{matrel}
\end{eqnarray}
In Eq.~(\ref{matrel}), $g_3$ is the QCD running coupling constant, $h_t$
and $h_b$ are the top and bottom Yukawa couplings, and the barred
quantities (e.g. $\bar{A}_t$) are the usual MSSM parameters divided by
the SUSY scale $M_S$.  The quantity 
$\tilde{a}\equiv\bar A_t - \bar\mu/\tan\beta$
and $t=\ln(M_S^2/m_t^2)$.
Only the leading terms in powers of $h_b$ and $\tan\beta$ have been retained,
and the small, ${\cal O}(h_t^2 M_Z^2)$ correction
to ${\cal M}_{12}^2$ has been explicitly included. 
Note, the above expressions hold only in the limit of small splittings
between the running stop masses, and the
condition $2m_t\max(|A_t|,|\mu|)<M^2_S$ must be fulfilled, with a
similar condition in the sbottom sector.
The analytical expressions 
presented above
are very useful, since they provide an understanding
of the behavior of the Higgs masses and mixing angles with the
stop and sbottom soft supersymmetry breaking masses 
and mixing parameters, and, unless the stop and sbottom mass
splittings are very large, they provide an excellent approximation
to the precise Higgs mass matrix elements.
However, the final, numerical results of our analysis make use of 
the complete one--loop RG improved effective
potential computation~\cite{CAESQUWA} of the Higgs
squared mass matrix elements, which allows a more reliable treatment
of the cases in which the squark mixing terms or the squark
mass splittings become large.

The approximation that $\sin\beta\simeq 1$, which is good
for moderate or large values of $\tan\beta$, is 
equivalent to the relation $v_2^2 \gg v_1^2$.  In this limit,
$H_2$ is the Higgs doublet mainly responsible for electroweak
symmetry breaking and the mass of the Higgs $\phi_W$
is well approximated by $\sqrt{{\cal M}_{22}^2}$, where
the $m_A$ dependence is suppressed by the large
$\tan\beta$ factor. On the other hand,  the mixing
angle $\alpha$ can be determined from the expression
\begin{eqnarray}
\sin\alpha \cos\alpha = {{\cal M}_{12}^2 \over 
\sqrt{({\rm Tr}{\cal M}^2)^2-4\det{\cal M}^2}}.
\label{sin2alpha}
\end{eqnarray}
In the limit that ${\cal M}_{12}\to 0$, either $\sin\alpha$ or 
$\cos\alpha\to 0$.  In the case that ${\cal M}_{11}^2 >
{\cal M}_{22}^2$, it is $\sin\alpha$ that vanishes.  Otherwise,
it is $\cos\alpha$ that vanishes.
Because ${\cal M}_{11}^2 \simeq m_A^2\sin^2\beta \simeq m_A^2$, it
is $\sin\alpha$ that is suppressed when the off--diagonal elements
of the quadratic mass matrix are small and $m_A$ is large.  
In both cases the suppression of the BR$(\phi \rightarrow b\bar{b})$
affects the Higgs $\phi_W$.
Observe that the tree-level contribution to the
Higgs matrix element ${\cal M}^2_{12}$ is suppressed by
a $1/\tan\beta$ factor. 
This factor does not lead in general
to a suppression of the effective 
$\phi_W$ coupling to bottom quarks,
but compensates the $\tan\beta$ enhancement of $h_b$ to render
it Standard Model--like.
What is emphasized above is an additional
suppression, which only takes place when ${\cal M}^2_{12}$ is 
significantly smaller than the tree level value. In general, the
radiative corrections are very important and depend on the
sign and size of $\bar\mu \times \bar A_{t}$ and $\bar\mu\times \bar A_b$.
The possibility of such effects will no be apparent if one assumes
$\mu\simeq 0$ or $A_t,A_b\simeq 0$.

Up to this point, the discussion has made use of the tree--level (but
QCD corrected) relation between the Yukawa couplings and the quark masses.
However, for large values of $\tan\beta$, there can be a significant
modification of the bottom and possibly tau Yukawa couplings from SUSY 
corrections~\cite{sola,CMW}.  In fact,
it is possible to enhance the bottom or tau coupling of the Higgs
boson 
independently of each other.
For completeness, we provide the modifications to the $\phi b\bar b$ 
couplings derived by us earlier~\footnote{These results agree with those obtained by 
diagrammatic computations\cite{dabelstein,susy99talks}.} in an effective Lagrangian approach
\cite{CMW}.
The starting point is the effective Lagrangian at energies
below the supersymmetric particle masses, which are assumed
to be larger than the weak scale, $M_S^2 \gg M_Z^2$,
\begin{equation}
{\cal L} \simeq h_b H_1^0 b \bar{b} + \Delta h_b H_2^0 b \bar{b}.
\label{couplings}
\end{equation} 
In the above,
the appearance of the one-loop suppressed coupling $\Delta h_b$ 
is a reflection of the breakdown of supersymmetry at low energies. 
The
CP-even Higgs boson
couplings to bottom quarks are approximately given by~\cite{CMW}
\begin{equation}
{h}_{b,h}  \simeq -\frac{m_b \sin\alpha}{v \cos\beta}\
\left[ 1 - \frac{\Delta(m_b)}{1 + \Delta(m_b)} \left( 1 + \frac{1}
{\tan\alpha \tan\beta} \right) \right],
\label{barhb}
\end{equation}
\begin{equation}
{h}_{b,H} \simeq \frac{m_b \cos\alpha}{v \cos\beta}
\left[ 1 - \frac{\Delta(m_b)}{1 + \Delta(m_b)} 
\left( 1 - \frac{\tan\alpha}{\tan\beta} \right) \right],
\label{tildehb}
\end{equation}
~\\
where ${h}_{b,h}$ and ${h}_{b,H}$ denote the couplings of the lightest
and heaviest CP-even Higgs boson respectively and $\Delta(m_b) =
(\Delta h_b/h_b) \tan\beta$.  
The coupling of the CP-odd Higgs boson is much simpler, and takes the form
\begin{equation}
h_{b,A} = 
{m_b\over (1 + \Delta(m_b)) v}\tan\beta \simeq h_b.
\label{bhCP}
\end{equation}

The function $\Delta(m_b)$ contains two main
contributions, one from a bottom squark--gluino loop 
(depending on the two bottom squark masses $M_{\tilde b_1}$
and $M_{\tilde b_2}$ and the gluino mass $M_{\tilde g}$) and another one
from a
top squark--higgsino loop (depending on the two top squark masses
$M_{\tilde t_1}$ and $M_{\tilde t_2}$ and the higgsino mass parameter
$\mu$).  The explicit form of $\Delta(m_b)$ at one--loop can be approximated
by computing the supersymmetric loop diagrams at zero external momentum
($M_S \gg m_b$) and is given 
by~\cite{deltamb,deltamb1,deltamb2,deltamb3}:
\begin{center}
\begin{eqnarray}
\Delta(m_b) \simeq {2\alpha_3 \over 3\pi} M_{\tilde g}
\mu\tan\beta~I(M_{\tilde b_1},
M_{\tilde b_2},M_{\tilde g})
 + {Y_t \over 4\pi} A_t\mu\tan\beta~I(M_{\tilde t_1},M_{\tilde t_2},\mu),
\label{deltamb}
\end{eqnarray}
\end{center}
where $\alpha_3=g_3^2/4\pi$, 
$Y_t= \frac{h_t^2}{4\pi}$, 
and the function $I$ is given by,
\begin{center}
\begin{eqnarray}
I(a,b,c) = {a^2b^2\ln(a^2/b^2)+b^2c^2\ln(b^2/c^2)+c^2a^2\ln(c^2/a^2) \over
(a^2-b^2)(b^2-c^2)(a^2-c^2)},
\end{eqnarray}
\end{center}
which is positive by definition.
Smaller contributions to $\Delta(m_b)$~\cite{deltamb2} 
have been neglected for the
purpose of this discussion.
The value of $\Delta(m_b)$ 
in Eq. (\ref{deltamb}) is defined
at the scale $M_{S}$,
where the sparticles are decoupled. 
The $ h_b$ and $\Delta h_b$  couplings  
should be computed at that scale, and run down with their respective 
renormalization group equations to the scale $m_A$, where the 
relations between the couplings of the bottom quark  to the neutral Higgs 
bosons and the running bottom quark mass are defined.

The CP-even Higgs couplings to the $\tau$-leptons are also affected
by large corrections at large $\tan\beta$. They are given 
by similar expressions as the ones for the bottom couplings, but
replacing $\Delta(m_b)$ by
\begin{eqnarray}
\Delta(m_\tau) \simeq {g_1^2 \over 16\pi^2} 
M_1\mu\tan\beta I(M_{\tilde\tau_1},
M_{\tilde\tau_2},M_1) + {g_2^2 \over 16\pi^2} M_2\mu\tan\beta \;
I(M_{\tilde\nu_\tau},M_2,\mu),
\end{eqnarray}
where $g_1$ and $g_2$ are the $U(1)$ hypercharge and $SU(2)$ weak isospin couplings.
Since it is proportional to weak couplings, $\Delta(m_\tau)$ is usually
much smaller than $\Delta(m_b)$; the exact value of $\Delta(m_\tau)$
depending on the relative size of the weak gaugino masses.

In the earlier discussion of the suppression of the $\phi b\bar b$
coupling, an implicit assumption was made that $\Delta(m_b)$ 
and $\Delta(m_\tau)$ were small.
Indeed, from Eq. (\ref{barhb}) 
(Eq. (\ref{tildehb})), we observe that, in the
limit $\sin\alpha = 0$ ($\cos\alpha = 0$), the $\phi b\bar b$ coupling
is given by 
\begin{equation}
{h}_{b,h} ({h}_{b,H}) = \frac{m_b}{\sin\beta v} \times \frac{\Delta(m_b)}
{(1 + \Delta(m_b))}  \equiv \Delta h_b,
\label{barhb2}
\end{equation}
which vanishes if $|\Delta(m_b)| \ll 1$. 
A similar expression to 
Eq.~(\ref{barhb2}) holds for the $\tau$ lepton coupling.

The bottom mass
correction factor $\Delta(m_b) \ll 1$,
except when $\tan\beta$ and/or 
the stop mixing mass parameters $A_t$ and $\mu$ are large,
in which case it can be near unity. 
When $\Delta(m_b)$ is of order 1, the 
$\phi b\bar b$ coupling can be of the order of the 
Standard Model one even though  $\sin\alpha\cos\alpha\to 0$.
Moreover, since $\Delta(m_b)$ and $\Delta(m_\tau)$ will be different in
general, their relative strength can be quite different from that in
the SM or the tree--level MSSM.  It is possible for $h_{\tau,h}$ to vanish,
while $h_{b,h}$ is substantial.  Note, 
a strong suppression of the bottom 
coupling ${h}_{b,h}$ can still occur for slightly different 
values of the Higgs
mixing angle $\alpha$, namely
\begin{equation}
\tan\alpha \simeq \frac{\Delta (m_b)}{\tan\beta} \equiv 
\frac{\Delta h_b}{h_b}.
\end{equation}

Under these conditions,
\begin{equation}
{h}_{\tau,h} = \frac{m_{\tau}}{v \sin\beta} \left( 
\frac{ \Delta(m_{\tau}) - \Delta(m_b)}{ 1 + \Delta(m_{\tau})} \right),
~~~~~h_{b,h}=0.
\label{nob_yestau}
\end{equation}
A similar expression is obtained for the coupling ${h}_{\tau,H}$
in the case ${h}_{b,H} = 0$.
Hence, if $\tan\beta$ is very large and $\Delta(m_b)$ is of order
one, the $\tau$ Yukawa coupling may {\it not} be strongly suppressed with
respect to the Standard Model case and can provide the {\it dominant}
decay mode for a Standard Model--like Higgs boson.  Likewise,
a suppression of the $h_{\tau,h}$ coupling arises for
$\tan\alpha\simeq\Delta(m_\tau)/\tan\beta$. 

At tree-level, when $\phi\tbb$ vanishes, so does $\phi\to\tau^+\tau^-$. 
Therefore,  the Higgs decays to $gg, c\bar c, W^*W^*$
and $\gamma\gamma$ occur at enhanced rates compared to the SM expectations.
Once the vertex corrections are included, the results will depend on
$\Delta (m_b)$ (we assume for the rest of this discussion that
$\Delta(m_\tau)$ is small). Since the $\Delta(m_b)$ 
corrections depend strongly on the 
size and sign of $M_{\tilde{g}}$, and hence introduce a dependence 
on parameters which do not otherwise affect the Higgs masses and 
mixing angles in
a relevant way, we shall neglect
them in the main analysis. However, in Section~\ref{sec:results}, 
we shall present a dedicated analysis of the possible effects of 
these corrections on the Higgs phenomenology.  

The vanishing of the $\phi b\bar b$ coupling may be problematic for
Higgs searches at LEP and the Tevatron.
The enhanced decays $\phi\to gg$ and $\phi\to c\bar c$ are difficult to observe
at the LEP 
and particularly at the Tevatron collider because of
increased backgrounds.  
The ``trilepton'' signature from
$W\phi(\to W^*W^*)$ may be challenging at the Tevatron \cite{JWHBMSAT},
because of the small signal rate.
While the cross section for the 
process $gg\to\phi\tgg$ can be enhanced up to about 10 fb at
the Tevatron collider, which may be observable, a detailed study of
the $\gamma\gamma$ backgrounds in the mass range around 100 GeV is 
still lacking.

We now turn our attention to the case of the LHC.
Search strategies in the $\gamma\gamma+X$ final state change from
the low luminosity run (collecting up to 30$~\ifb$) to the high luminosity
run (30 to 100$~\ifb$ and up to 300$~\ifb$) \footnote{We use the
term luminosity interchangeably to mean instantaneous luminosity
and total integrated luminosity.  The meaning should be clear from
context.} because of the relative behavior of the signals and backgrounds. 
At low luminosity,
the experiments are most sensitive to the subprocess $gg\to\phi\tgg$.  Given
the reach for a SM Higgs boson, the reach in the MSSM can
be calculated using the factor $R^{'}(m_\phi)$:
\bea
\label{ratiorp}
R^{'}(m_\phi)={\Gamma(\phi\to gg)\; {\rm BR}(\phi\tgg)
\over \Gamma(\phi\to gg)_{SM}\; {\rm BR}(\phi\tgg)_{SM}}.
\eea
At high luminosity, the best reach 
for a Higgs with SM-like couplings to the gauge
bosons is in the $W\phi_W(\tgg)$ and
$t\bar t\phi_W(\tgg)$ channels.  In this case, the production cross
section depends on a tree level coupling, and loop effects arise
only in BR$(\phi_W\tgg)$.  
The relevant factor
$R^{''}(m_\phi)$ is
\bea
\label{ratiorpp}
R^{''}(m_\phi) & = & \left\{\s2bma,\c2bma\right\}  
{{\rm BR}(\phi\tgg)\over {\rm BR}(\phi\tgg)_{SM}}  \nonumber \\
& \simeq & \left\{\cos^2\alpha,\sin^2\alpha\right\}
{{\rm BR}(\phi\tgg)\over {\rm BR}(\phi\tgg)_{SM}}. 
\eea

Let us analyze the properties of the Higgs sector relevant to $R^{'}$
and $R^{''}$ in more detail.
The effect of a light top squark (or light bottom squark at large
$\tan\beta$) with $\tilde A_t$ ($\tilde A_b\equiv A_b-\mu\tan\beta$) large
is to decrease the top quark contribution 
to these
loop effects.  This decreases the partial width $\Gamma(\phi_W\to gg)$,
but increases BR$(\phi_W\to\gamma\gamma)$ since in the Standard Model
the $W$ and $t$ contributions destructively
interfere, with the former being the dominant one.
It was noted earlier that there is a reciprocal relation between
$\Gamma(\phi_W\to gg)$ and $\Gamma(\phi_W\tgg)$ because of the domination
of the $WW$ loop in the latter \cite{DJOUADI}.  
Because BR$(\phi_W\tgg)=\Gamma(\phi_W\tgg)/\Gamma_{tot}$, the relation is
not entirely compensating, since BR$(\phi_W \tgg)$ depends on
$\Gamma_{tot}$ and hence on the width of the Higgs decay into
bottom quarks and tau leptons.
For small values of $\tan\beta$ or
large values, the factor $R^{'}$, Eq. (\ref{ratiorp}) can be significantly
decreased \cite{DJOUADI} because of a cancellation between
the top quark loops and the stop and sbottom loops.  
Of course, the presence of light sparticles implies that
the next generation of experiments can directly probe them.
The advantage of the $t\bar t\phi_W$ channel is that $R^{''}$
depends on BR($\phi_W\tgg)$) and not $\Gamma(\phi_W\to gg)$.
Therefore, the decrease in $R^{'}$ can be compensated by an
increase in $R^{''}$.

Another way to modify $R^{'}$ and $R^{''}$ is to change one of the
tree level couplings listed above.  
For instance, when $\tan\beta$ is large and
$\sin\alpha$ is small, the lightest CP-even Higgs
boson presents Standard Model--like couplings to the $W$ and 
$Z$ gauge bosons.
However, even when $\sin\alpha$ is small, the product
$\sin\alpha \times \tan\beta$ is not determined a priori, and
depends on the exact characteristics of the supersymmetric spectrum.
When $\sin\alpha \times \tan\beta$ is larger than one, the
BR$(h\to b\bar{b})$ becomes larger than the SM one, and, since it
is the dominant Higgs decay channel, it partially suppress the
BR$(\phi \to \gamma\gamma)$. On the contrary, if BR$(h\to b\bar{b})$
becomes much smaller than one, something that, as we explained
above,  can happen in
certain regions of parameters~\cite{CMW}, the BR$(\phi \to \gamma\gamma)$
will be strongly enhanced, improving the LHC prospects of finding
a light CP-even Higgs without going to the highest luminosity runs. 

Our discussion of searches for a SM--like Higgs boson at the LHC
is limited to those presented by the experiments themselves.
{\sc ATLAS} has presented encouraging numbers for their reach in
the $t\bar t\phi(\tbb)$ channel.\footnote{At low luminosity,
the process $W\phi(\tbb)$ can also be used.}  In the
approximation of Table I, we observe that this signature has large or
small couplings in the same regions of MSSM parameters space as for
$V\phi(\tbb)$ at the LEP and Tevatron colliders. 
This 
potentially important channel is not included in our analysis. 
Let us mention, however, that, if the Higgs were discovered at LEP and/or
the Tevatron, this channel will serve to measure its 
coupling to the top quark, providing an important independent test
of the origin of fermion masses. On the other 
hand, if the effective luminosity at the 
next run of the Tevatron were small, 
this channel will provide additional means for the LHC to test the Higgs 
responsible for electroweak symmetry breaking in most of the MSSM parameter
space. Hence, this Higgs production channel at LHC adds to the
complementary physics potential of the three colliders that we
are emphasizing in this article. 

\section{Results}
\label{sec:results}

For our analysis, we rely on the projected discovery reach of the
three colliders for a Standard Model Higgs boson.
For LEP2, running at $\sqrt{s}=200$ GeV and collecting
200 pb$^{-1}$ of data (per experiment), we use the numbers of
Ref.~\cite{janot}.  For the Tevatron in Run II and Run III, we use the
results of the Higgs Working Group of the Workshop
on Physics at Run II~\cite{SHWG}.  
For the
LHC, we use the technical design reports and updates of the
{\sc ATLAS}\cite{RICHTER} and {\sc CMS}\cite{CMS} collaborations.
A comparison of the sensitivity of the two experiments in the 
$\gamma\gamma$ channels reveals that {\sc CMS} is substantially
more sensitive than {\sc ATLAS}.  Therefore, only the projected
reach of the {\sc CMS} detector is explicitly used in our analysis
to represent the reach of the LHC.
To demonstrate the potential of the LHC if 300$~\ifb$ of data is 
collected, we scale the 100$~\ifb$ significances by a factor of $\sqrt{3}$.
The significance in the MSSM is determined by rescaling the partial widths
and branching ratios accordingly.

As discussed in Sec.~\ref{sec:couplings}, some choices for the
soft supersymmetry-breaking parameters at the weak scale can make 
it difficult to observe the Higgs $\phi_W$ at LEP and the
Tevatron in the $\phi_W\to b\bar b$ channel or LHC in the 
$\phi_W\to\gamma\gamma$
channel.  In this section, we demonstrate that, quite generally, 
difficulties will not arise
in {\it both} channels, provided that the 
experiments operate efficiently and
receive enough integrated luminosity.  

We construct our argument by concentrating on MSSM parameters that
are problematic for one of the channels, and then displaying the
complementarity of the other channel.  Our choices include the
possibility of light top and bottom squarks, which can have an important
impact on one-loop suppressed decays of $\phi_W$.
Higgs boson properties are calculated using {\sc HDECAY}~\cite{hdecay}.
However, due to the relevance of the bottom mass corrections in
defining the proper bottom quark Yukawa coupling at the scale $M_S$,
we have modified the effective potential calculation of 
Ref.~\cite{CAESQUWA}, used in {\sc HDECAY},
in order to include these corrections in the large $\tan\beta$ regime.
The inclusion of these corrections in the expression
of the Higgs masses lead to non-leading logarithmic
two-loop corrections. Although in a complete two-loop calculation
there might be other non-logarithmic 
terms modifying the Higgs masses, these corrections
are particularly important, since they can modify the one-loop result
by factors of order one.

\subsection{The Minimal Mixing Model}
\label{subsec:minmod}

The case when $\tilde{A}_t=0$ yields the smallest value for the
upper bound on the Higgs boson mass for a given choice of
$\mu$ and stop masses (for large values of $\tan\beta$,
small sbottom masses or large values of $\mu$ tend to lead to
even lower values for the upper bound~\cite{CAESQUWA}).  
We fix the overall scale of supersymmetry particle masses  $M_S=1$ TeV, 
with $M_S^2 = \frac{1}{2}(m_{{\tilde t}_1}^2+m_{{\tilde t}_2}^2)$.
Furthermore, for comparison with the results presented by {\sc ATLAS}
and {\sc CMS}, 
we choose $\mu=-100$ GeV, which is small enough that it does
not differ significantly from the $\mu =0$ case, but large
enough to avoid a light chargino.  In our scan of the $m_A-\tan\beta$
plane, we tune $A_t=\mu/\tan\beta$ to achieve the conditions of
minimal mixing at each point in the plane.
This choice of parameters yields a moderate upper bound for the 
lightest Higgs boson mass which is, however, beyond the kinematic reach of LEP.
This is a conservative assumption for the LHC experiments,
since the discovery reach is typically best for
a Higgs $\phi_W$ mass near the highest allowed upper bound in the MSSM.

The regions of the $M_A-\tan\beta$ plane
in which a Higgs $\phi_W$ boson would be discovered at the
5--$\sigma$ level at 
the Tevatron are shown as shaded regions in Fig.~\ref{minimal}(a). 
Different
shadings correspond to different assumptions about the luminosity
delivered to {\it both} experiments.  The LEP discovery contour
(for a Higgs $\phi_W$ boson) is shown by the double line (the region
below the contour will be probed).  With
{\it more} than 5$~\ifb$, the Tevatron begins to provide information
beyond that from LEP, provided that the Higgs boson is not
already discovered.  Up to 30$~\ifb$ of data is needed to cover the
problematic region around $\s2bma\simeq\c2bma$ at low $\tan\beta$
where $\mh$ and $\mH$ are separated by more than
$\simeq 10$ GeV~\cite{Baer,CMW}.

In Fig.~\ref{minimal}(b), the corresponding discovery 
reach with the {\sc CMS} experiment
is shown.  The complementarity
of the colliders is clear.  The LHC experiments are most sensitive to
large $m_A$, where $\mh$ approaches its upper bound for the given
choice of parameters..  As $m_A$ decreases, the 
$\phi b\bar b$ coupling increases, so that $B(\phi\tgg)$ and $R^{'}$
and $R^{''}$ decrease.
At the Tevatron, the region of large $m_A$ and large
$\tan\beta$ is harder to cover, 
because  the $\phi b \bar b$ coupling decreases towards the
Standard Model value, yielding $R \simeq 1$.  Simultaneously,
the lightest CP-even Higgs boson mass approaches its
upper bound, so that more integrated luminosity is needed
to probe this region.
It is worth noting the change in the shape of the LHC contours
between low and high luminosity, when the $gg\to\phi$ process 
becomes less important than the $t\bar t\phi/W\phi$ process.
After the low luminosity run at the LHC (30~$\ifb$),
if the Tevatron obtains only 10~$\ifb$ of data, 
there remains a region uncovered by both colliders for
$m_A\simeq 250-300$ GeV (there
would also still be the hole at $\s2bma\simeq\c2bma$).  The
high luminosity run (but only with $100~\ifb$)
is necessary to guarantee full coverage in
this region,
unless the Tevatron collects 20~$\ifb$ or more of 
integrated luminosity per experiment.

\subsection{$\phi b\bar b$ suppression for large $M_S$}

Figure~\ref{mix_large_ms} was generated using the choice
of MSSM parameters $A_t=-\mu=1.5$ TeV 
and $M_S=1$ TeV, which
yields a suppression of the Higgs boson coupling to $b\bar b$ 
(and $\tau^+\tau^-$) in a significant portion of the $m_A-\tan\beta$
plane
for the same Higgs boson that couples strongly to $W$ and $Z$ bosons 
($\phi_W$).  
An approximate analytic formula that shows the necessary relations between
parameters can be obtained by setting ${\cal M}_{12}^2 = 0$ in 
Eq.~(\ref{matrel})
and  was presented in Ref. \cite{CMW}.
In Fig.~\ref{mix_large_ms}(a), the vanishing of $\sin\alpha$ is seen when
$h \equiv \phi_W$ (region  which remains uncovered by
the Tevatron, in the upper part
of the Figure),
whereas the vanishing
of $\cos\alpha$ occurs when $H \equiv \phi_W$ (region which remains
uncovered by the Tevatron, in
the lower part of the Figure ). 
As alluded to earlier,
the vanishing of the $\phi_W b\bar b$ and $\phi_W\tau^+\tau^-$ 
couplings greatly
enhances BR$(\phi\tgg)$, and {\sc CMS} has 
little
difficulty in covering the complementary regions in the low luminosity run,
as shown in Fig.~\ref{mix_large_ms}(b).

After combining both discovery reaches, a small region, for which
$\s2bma\simeq\c2bma$ persists, however, uncovered
by both colliders.
For clarity, we wish to emphasize again that LHC will have other means
of testing the region of parameters which remain uncovered in our
analysis by means of the production of other Higgs bosons of the low
energy effective theory~\cite{RICHTER}.
However, in this region of parameters, 
none of these Higgs bosons will have
SM--like couplings, and hence will not directly test the mechanism
of electroweak symmetry breaking. In addition,
one of the unambiguous predictions of weak-scale
supersymmetry (the upper bound on the mass of the SM--like Higgs) will
not be tested directly.  

\subsection{Bottom Yukawa corrections}
\label{subsec:bottaucorr}

So far, our analysis has neglected corrections to 
the fermion Yukawa couplings from
loop corrections.  To illustrate the potential importance of 
these effects, we must
specify other parameters of the MSSM which, in the 
next--to--leading--logarithmic approximation,  do not 
affect the CP-even Higgs masses and mixing angles in a relevant way
(apart from the effects arising from the redefinition of the
$b$-Yukawa coupling, which we will discuss below).
Care must be taken in choosing these parameters, since an additional
enhancement of the $b$ Yukawa coupling 
at large values of $\tan\beta$ can cause the theory to become
non--perturbative.  Therefore, we consider an example similar to
the one presented in Fig.~\ref{mix_large_ms}, taking
${A_t}=-\mu= 1$ TeV.   This still illustrates the suppression of
the $\phi b\bar b$ coupling, but in a smaller 
region of the $M_A-\tan\beta$ plane.   
For this same choice of parameters,
we then consider the effects of $\Delta(m_b)$
when the gluino mass parameter $M_{\tilde{g}}$ 
takes the values $\pm .5$ TeV.  
From Eq.~(\ref{deltamb}), we observe that the top
quark-Yukawa coupling induced-contribution is
negative for $A_t\mu<0$, while a positive (negative) gluino mass will 
decrease (increase) $\Delta(m_b)$.
The case of an overall positive correction (negative gluino mass) is
illustrated in Fig.~\ref{deltamb1_neg_mg}.  The effect of the correction
is minimal, since $\Delta(m_b)$ is never larger than about $0.3$ at large
$\tan\beta$.  For the case of a negative correction, however, as seen
in Fig.~\ref{deltamb1_pos_mg}, the suppression of the $\phi b\bar b$
coupling for $m_A \simgt 130$ GeV
is shifted to lower values of $\tan\beta$.  Note that the
suppression is now occurring because 
$\tan\alpha\simeq \Delta(m_b)/\tan\beta$,
and {\it not} because the tree-level
coupling vanishes $\sin\alpha\cos\alpha/\cos\beta\to 0$.
According to Eq.~(\ref{barhb2}), in the regions of parameter
space where the tree-level
$\phi b\bar b$ vanishes, the $\phi b\bar b$ coupling can be
substantial and the $\phi\tau^+\tau^-$ coupling is suppressed.
When the $\phi b\bar b$ coupling suppression occurs, the 
$\phi\tau^+\tau^-$ coupling will typically not vanish,
as demonstrated in Eq.~(\ref{nob_yestau}).

There are several phenomenological consequences of the mismatch
in the behavior of the $\phi b\bar b$ and $\phi\tau^+\tau^-$ couplings.
First, in the previous examples, we observed that the
$\phi_W\to b\bar b$ channel at the Tevatron (or the LHC)
can cover those regions where the $\phi_W\to\gamma\gamma$ channel
is suppressed.  
Secondly, in such examples, the simultaneous vanishing 
of the $\phi b\bar b$ {\it and} the $\phi\tau^+\tau^-$ couplings
led to an enhancement of BR($\phi\tgg$).  However, the enhancement
in the region where $\phi b \bar b$ is suppressed
will generally not be as large as naively expected, since 
$\Gamma(\phi\to\tau^+\tau^-)$ can still be substantial. Still,
the complementarity of the Tevatron and the LHC experiments in
the search for the Higgs $\phi_W$ remains clear.
Given
the fact that the decay $\phi_W\to\tau^+\tau^-$ may be dominant
in this region,
it is important to consider $\phi_W\to\tau^+\tau^-$ signatures at
the Tevatron and LHC. Preliminary results in this direction have
been presented in Ref.~\cite{wwfusion}.

An illustration of the possible variation of the $\phi_W 
\rightarrow b \bar{b}$ and $\phi_W \rightarrow \tau^{+}
\tau^{-}$ decay modes in parameter space 
is presented in Fig.~\ref{tau_vs_bbbar},
which shows the function $R$ of Eq.~(\ref{rone}) for the $b\bar b$
final state ($R_b$)
(a) and the $\tau^+\tau^-$ final state ($R_\tau$)(b)
for the same parameters as in Fig.~\ref{deltamb1_pos_mg}.  Note the large
region where $R_b$ is greater than the SM value, where as $R_\tau$
is below it.  Of course, there is a compensating region along contours
of $\tan\alpha\simeq\Delta(m_b)/\tan\beta$ at very large $\tan\beta$
where $\phi_W\to\tau^+\tau^-$ is the dominant decay mode.  In terms of
area covered in the $M_A-\tan\beta$ plane, the former is more significant
than the latter.  It is also worth noting that both $R$ values approach
their SM values in the limit of large $M_A$.  Indeed, when $M_A$ is
large {\it and} $M_S$ is large, the effect of the radiative corrections
vanish.  When $M_A$ is fixed and $M_S$ is arbitrarily large, however, the 
radiative corrections will in general be relevant~\cite{CMW}, a
reflection of the lack of supersymmetry in the low energy effective
theory.

It is important to emphasize that the nice complementarity between
the $\phi_W \to b\bar{b}$ and $\phi_W \to \gamma\gamma$ 
decay channels holds due to the fact that the former is, in general,
the dominant Higgs decay channel. This complementarity does not
extend to the $\phi_W \to \tau^+\tau^-$ channel. Indeed, in the
above example, we see region of parameters for which $\phi_W \to
\tau^+\tau^-$ is suppressed and, due to a slight enhancement of
$\phi \to b \bar b$ decay rate in the same region of parameters,
the LHC reach in the $\phi_W \to \gamma \gamma$ channel is also
suppressed.

\subsection{Cancellations from the sbottom sector}

In the previous examples, we showed how relatively large values of 
${A}_t\simeq -\mu$ led to the suppression of the $\phi b\bar b$
coupling in the large
$\tan\beta$ regime.  We demonstrate that this can also occur
when $A_b\simeq \mu$ in Figs.~\ref{ab_eq_mu_pos_mg},
and~\ref{ab_eq_mu_neg_mg}. The effect of $A_b$, however, is in
general much weaker than the $A_t$ one, and becomes only
relevant for large values of the bottom Yukawa coupling, that is
for values of $\tan\beta \simeq m_t/m_b$. To show this, in
these figures, we choose moderate values of
$A_b $ and $\mu$, taking $A_t = 0$. The vanishing of $A_t$ 
leads, in general, to masses that are of the order of the
ones obtained in the minimal mixing case, although they
can be further reduced, due to the $\mu$--induced terms 
discussed above, for large values of $\tan\beta$. 
Also, the top Yukawa contribution to $\Delta(m_b)$ vanishes
for $A_t=0$.
In Fig.~\ref{ab_eq_mu_pos_mg}, 
we have chosen $A_b=\mu=1.25$ TeV and
$M_{\tilde g}=.5$ TeV, leading to $\Delta(m_b) >0$. No relevant
modification in the reach is found compared to the minimal mixing case
(Fig.~\ref{minimal}).
Indeed, for the
parameters chosen, the suppression of 
the $\phi b\bar b$ coupling takes place at values of $\tan\beta$
larger than the ones considered in this analysis.
To observe the suppression for  positive corrections, 
we would need to choose larger values of $A_b$ and $\mu$.

For $M_{\tilde g}=-.5$ TeV, which yields $\Delta(m_b)<0$, 
we observe the suppression 
of the $\phi b\bar b$ coupling in Fig.~\ref{ab_eq_mu_neg_mg}, in 
region of parameter space in which
$\tan\alpha\simeq\Delta(m_b)/\tan\beta$.  This leads to holes in
the regions of parameter space to be tested at LEP and the Tevatron
in the
$\phi_W\to b\bar b$ channel, but a compensating increase in
$\phi_W\to\gamma\gamma$ rate in
the same regions of parameters.  Since we have also included 
the bottom Yukawa
coupling corrections to the calculation of the mass spectrum, we observe
the reappearance of the LEP exclusion at large $\tan\beta$, due to
a reduction of the Higgs boson mass.

Although the above provides only an extreme example for which
only the $A_b$-induced effects were considered, it shows in 
a clear way the relative importance of the effects on the
Higgs mass matrix elements induced by the presence of non-trivial
mixing in the sbottom and stop sectors. In the following sections, 
we shall present examples in which $A_t \simeq A_b \neq 0$.

\subsection{Light stop and large mixing}

In the minimal mixing model investigated in section~\ref{subsec:minmod}, 
the SUSY 
scale $M_S$ was set to 1 TeV,
but the mixing term $\tilde{A}_t$ was zero.  In the present case, we consider
equal values for the left- and right-handed soft supersymmetry-breaking
squark masses of the third generation, and adjust the common value 
to yield a lightest top squark mass of $200$ GeV.
This is meant to demonstrate the possibility of large corrections to the
$\phi gg$ and $\phi\gamma\gamma$ couplings
from light sparticles.  In addition, this lowers the scale $M_S$.
We fix $\mu=-.3$ TeV, and $A_t=1$ TeV, so that stop mixing is 
large~\footnote{As has been
observed in Ref.~\cite{DJOUADI}, precision electroweak measurements 
tend to disfavor the presence of light third generation squarks with
large mixing angles. Most of the parameters considered in this subsection,
as in the following two, are 
only marginally consistent with the precision
electroweak data.}.  
This is motivated by the form of the 
$\{h,H\}\tilde{t}\tilde{t}$ coupling (written
here in the interaction basis):
\begin{eqnarray*}
{ 2 m_t^2 \over v \sin\beta } 
\left\{\cos\alpha,\sin\alpha\right\} 
- {2 M_Z^2 \over v}
\left\{\sin(\alpha+\beta),-\cos(\alpha+\beta)\right\} 
(1/2-2/3\sin^2\theta_W) &~~~(LL)\\ 
\nonumber
{2 m_t^2  \over v \sin\beta } 
\left\{\cos\alpha,\sin\alpha\right\} - {2 M_Z^2 \over v}
\left\{\sin(\alpha+\beta),
-\cos(\alpha+\beta)\right\} 2/3\sin^2\theta_W 
&~~~(RR)  \\ 
\nonumber
\left.{m_t  \over v \sin\beta } 
\left[ A_t \left\{\cos\alpha,\sin\alpha\right\} + 
\mu \left\{\sin\alpha,-\cos\alpha\right\}\right] \right.&~~~(LR)
\end{eqnarray*}
where we have
denoted the components in parentheses.
For large $LR$ mixing, the terms proportional to 
$A_t$ and $\mu$ can dominate the stop-Higgs couplings.
For a Higgs with SM-like couplings to the gauge
bosons, in the moderate and large $\tan\beta$ regime,
$A_t$ is the relevant mixing parameter determining
the strength of this coupling. For large values
of $m_A$ and arbitrary values of $\tan\beta$, this
coupling is proportional to $\tilde{A}_t$, which
is approximately equal to $A_t$ in the 
large $\tan\beta$ regime.

In this region of parameters, the Tevatron can discover 
a Higgs $\phi_W$ in most
of the parameter space with about 20$~\ifb$.
One observes from Fig.~\ref{light_stop}(a) that 
although the stops are lighter in this case, the
reach at the Tevatron is somewhat
suppressed with respect to the minimal
mixing case, and 
30$~\ifb$ are necessary to cover the whole parameter space, with 
the exception of the region of parameters for which
$\s2bma\simeq\c2bma$.  The origin of the relative
suppression in the discovery reach is  the upper bound on the
lightest CP-even Higgs mass, which increases
substantially when $\tilde{A}_t\ne 0$. For instance, while for
the minimal mixing case this upper bound is below
115 GeV, values close to 120 GeV are obtained in the case under
analysis.

As noted earlier, light top and bottom squarks can have a large effect on
the one--loop suppressed partial widths $\Gamma(\phi\to gg)$ 
and $\Gamma(\phi\tgg)$.
The relation between these two quantities is reciprocal, and depends on the
size of $A_t$ and $\mu$.  When the stop mixing mass
parameters (in particular $A_t$) become large, the width
$\Gamma(\phi\to gg)$ can be greatly decreased since the light top 
squark loop contribution can partially or totally
cancel the top quark loop induced one.  
In Fig.~\ref{light_stop}(b), it is clear that $R^{'}$, 
Eq. (\ref{ratiorp}) 
is widely suppressed,
and this is reflected in the fact  that the low luminosity run 
of the LHC cannot 
observe the Higgs boson that
couples strongly to the $W$ and $Z$ boson
in the gluon fusion channel. In the high luminosity run,
the dependence on $\Gamma(\phi\to gg)$ is weakened, and the
reach is dominated by the $t\bar{t}\phi/W\phi$ channels (see 
also~\cite{stophiggs}).  In this Higgs mass range, the LHC sensitivity
in these channels depends only weakly on  the Higgs mass and
for high luminosity the LHC discovery reach becomes
similar  to the one in the minimal mixing case.

The complementarity of the Tevatron and LHC colliders in this case
is similar to the case of minimal mixing, although somewhat
larger luminosities in both colliders
are needed in order to obtain full coverage
of the MSSM parameter space.

\subsection{$\phi b\bar b$ coupling suppression for light stops}

Since we have identified two interesting effects, namely the suppression
of the $\phi b\bar b$ coupling from Higgs mixing and the suppression of
one--loop couplings from large stop mixing, one may wonder if both can
occur.  The conditions for the cancellation of ${\cal M}_{12}^2$, and hence
for $\phi b\bar b$ suppression, can be determined analogously to the large
$M_S$ case, but since the stop mixing effects
become larger, the approximate formulae Eq. (\ref{matrel}) is no longer
valid, and one should work with the full effective potential
computation~\cite{CAESQUWA}.  One example occurs for $\mu=1$ TeV,
$A_t=.65$ TeV, and a lightest stop mass fixed at 200 GeV as before.
We display an example in 
Figure~\ref{mix_small_ms}.
Since the stop mixing effects we are considering are larger than in
the previous cases, the cancellation of the $\phi b \bar b$ coupling
occurs for larger values of the CP-odd Higgs mass. Indeed,
Fig.~\ref{mix_small_ms}(a) reveals that the region of
$\phi b\bar b$ suppression is shifted substantially.  

Most interesting is
the fact that LEP could discover a Higgs boson with strong couplings to
$W$ and $Z$ bosons at large $\tan\beta$.  As noticed before, this
is just a reflection of the suppression of the Higgs mass induced
by the $\mu$ parameter at large values of $\tan\beta$, as shown in
the expression of ${\cal M}_{22}^2$ in Eq.~(\ref{matrel}). 
Of course, in the regions of $\phi b\bar b$ suppression, neither LEP nor
the Tevatron have any reach (we are not considering a possible enhancement
of the tau lepton coupling, which, as explained before, can take place
if $\Delta(m_b) \neq \Delta(m_\tau)$). 

Because of the suppression of the $\phi b\bar b$ coupling, the LHC,
Fig.~\ref{mix_small_ms}(b), has significant reach in the regions 
complementary to LEP and the Tevatron.  Observe that the reach at low
$\tan\beta$ and large $m_A \simgt 350$ GeV
is slightly less than for the minimal mixing model, because
of a small increase of the $\phi b\bar b$ coupling. However, for
the same value of $\tan\beta$ and $m_A$, the lightest CP-even
Higgs mass increases, and therefore the increase of the
BR($h \rightarrow b \bar{b}$) is not reflected in an increase of
the Tevatron reach.

More important, there are region of parameters, at large values of
$\tan\beta$,  for which the
Higgs becomes accessible to the three colliders with relatively
small luminosity. This would provide a perfect situation: A
Higgs with SM-like couplings to the gauge bosons
will be discovered at LEP by the end of the year 2000, and
its properties will be further tested at the Tevatron and LHC 
colliders.

\subsection{$\phi b \bar{b}$ coupling enhancement and light
third generation squarks.}

A large value of $\tilde{A}_b\equiv A_b - \mu\tan\beta$ can have
significant consequences when the bottom squark is also light.
In Fig.~\ref{light_sbot}(b), one observes a general suppression of 
$R^{'}$ and $R^{''}$ throughout most of the $M_A-\tan\beta$ plane.
For this example, we have set $A_t=A_b=-0.5$ TeV,
$\mu=-1$ TeV, and all third generation squark parameters equal and
tuned to yield a bottom squark mass of $200$ GeV. For small 
values of the ratio of Higgs vacuum expectation values
$\tan\beta < 2)$ this choice of parameters may lead to too
light top squarks, with masses below the present experimental
bound, or even negative ones. 
In these cases, we have increase the squark masses
by setting a lower bound on
the lightest bottom squark of about 300 GeV. Although this
implies a slight discontinuity of the parameters chosen to 
perform the figures, this does not affect
the physical results since for such light stops and sbottoms
the $\tan\beta < 2$ regime is already ruled out by LEP2 data.
  
Since $\tilde A_t$
and $\tilde A_b$ are large, and the bottom and top squarks are light,
one might expect that $\Gamma(\phi\to gg)$ will decrease and
$\Gamma(\phi\tgg)$ will increase.  This is true, but any further
expectation that BR$(\phi\tgg)$ increases is nullified by the fact
that, for this choice of parameters, the bottom quark
coupling to the Higgs $\phi_W$ is enhanced with respect to the
Standard Model expectation, inducing a 
decrease of the BR$(\phi_W\tgg)$. Had we chosen the opposite sign
of $\mu$, an increase of the BR($\phi_W\to \gamma\gamma$) 
with respect to the SM value would have been observed, as in the previous
section.

Because of the enhancement of BR$(\phi\to b\bar b)$ and given that the
Higgs $\phi_W$ mass is below 112 GeV (it becomes smaller  at
small and large values of $\tan\beta$), it is particularly easy to probe
this region of the MSSM parameters. For instance, LEP can probe a 
significant portion of the
$M_A-\tan\beta$ plane (LEP will be sensitive to the region connected
by the narrow strip around $M_A\simeq 135$ GeV) as seen in 
Fig.~\ref{light_sbot}(a), and  the Tevatron would only
need a substantial luminosity upgrade to cover the difficult
region near $M_A\simeq 120$ GeV. 

\section{Conclusions}
\label{sec:conclusions}

To test if the mechanism of electroweak
symmetry breaking is indeed generated by a fundamental scalar, as
postulated in the minimal Standard Model and its supersymmetric
extension, 
it will be necessary to observe
the Higgs boson which is responsible for the $W$ and $Z$ boson masses, and
hence, has SM--like couplings to these gauge bosons.  
Otherwise, it will be difficult to conclude if the mechanism of
electroweak symmetry breaking relies on 
a weakly or strongly coupled model, or, if the low energy Higgs
sector fulfills the properties demanded in the MSSM, for instance
the upper bound on the lightest CP-even Higgs mass. In theories
with more than one Higgs doublet, the real part of the neutral
Higgs combination which acquires vacuum expectation value is
not necessarily associated with a physical mass eigenstate. However,
there are large regions of the MSSM parameter space where one of the
CP-even Higgs bosons couples in a Standard Model way to
the $W$ and $Z$ bosons and up quarks and, hence, can be identified as
the dominant source of electroweak symmetry breaking. We have denoted
such a Higgs boson as $\phi_W$.  Interestingly
enough, the $\phi_W$ couplings to bottom quarks and $\tau$-leptons,
which control the dominant decay modes of the Standard Model
Higgs, can be highly  non-standard. Therefore,
the full experimental program to discover the Higgs $\phi_W$
must account for the possibility that some of its standard signatures may be 
significantly modified. A suppression of
the standard signatures may
occur for natural choices of the soft 
supersymmetry-breaking
parameters of the low-energy effective theory.
When viewed
in this light, one sees that measurements at LEP and Tevatron can provide
important information about the Higgs sector which will be complemented by
measurements at the LHC.  On the other hand, in the regions of parameter
space where the Higgs searches at the LEP and Tevatron colliders become
difficult, the LHC will, in general, be able to find the Higgs 
$\phi_W$ at
relatively low integrated luminosities, ${\cal L} < 100$ fb$^{-1}$.
We have presented several explicit choices of 
MSSM parameters which demonstrate this point, including the possibility that
the top and bottom squarks are relatively heavy or light.  

In general, we have observed some patterns in the choices of soft supersymmetry-breaking
parameters that lead to difficulties at either LEP/Tevatron or LHC, but give
a complementary enhanced signature at the other collider(s):
\bit
\item When $A_t\mu<0$ or $A_b\mu>0$, with parameter values of the order of
the scale $M_S\simeq 1$~TeV, there can be a suppression of the 
$\phi_W b\bar b$ coupling,
which limits the $\phi_W\to b\bar b$ channel at LEP/Tevatron.  
Complementary to this, BR($\phi_W\tgg$) is enhanced at the LHC. Hence, 
while for these conditions the discovery of the Higgs $\phi_W$ at LEP and
the Tevatron will require very high luminosities, even a low luminosity
run at the LHC will be sufficient to discover the Higgs $\phi_W$.

\item As the values of $A_t, A_b$ and $\mu$ are lowered, 
the $\phi_W b\bar b$ 
coupling can still be suppressed in the presence of 
large radiative corrections to the $\phi_W b \bar{b}$ coupling, 
$\Delta(m_b)$, which are proportional to $\tan\beta$ and can
be of order one for large $\tan\beta$.
In this case, the suppression occurs because
the relation $\tan\alpha\simeq \Delta(m_b)/\tan\beta$ holds.  
There is also a mismatch between the $\phi_W b\bar b$
and $\phi_W\tau^+\tau^-$ couplings, to be discussed below.  
As before, the complementarity arises because the LHC reach
in the $\phi_W\tgg$ channel increases (decreases) when the bottom 
Yukawa coupling
is decreased (increased).

\item If $M_S$ is decreased, but the other parameters still obtain values
near 1 TeV, then the $\phi_W b\bar b$ suppression 
occurs for smaller values of
$\tan\beta$ and at large values of $m_A$.  At large $\tan\beta$, 
the mass of the Higgs $\phi_W$ decreases
from its upper bound, which is achieved at intermediate values
of $\tan\beta$ between 10 and 20, and LEP becomes sensitive to the Higgs $\phi_W$,
but not in the regions where
the Tevatron reach is also suppressed.  The signal rate 
in the $\phi_W\tgg$ channel
 is again enhanced in those regions inaccessible at the LEP/Tevatron.

\item A small top squark mass, a large value for $A_t$ and moderate
$\mu$ can decrease
$\Gamma(\phi_W\to gg)$ through the interference of top and top squark 
loops, which limits the channel $gg\to\phi_W\tgg$ at the LHC.
Simultaneously, the $\phi_W b\bar b$ coupling can be enhanced or suppressed
over the SM value,
because of the contribution of $A_t$ and $\mu$ 
to the mixing in the 
Higgs sector. If the $\phi_W b \bar b$ coupling is enhanced, the
BR$(\phi_W\tgg)$ can actually also decrease.  Because of the increase
in BR$(\phi_W\tbb)$, the channel $\phi_W\tbb$ at the Tevatron can be
used to cover the problematic
regions at the LHC, provided that the experiments at the
Tevatron receive enough luminosity. On the other hand, if the $\phi_W b \bar b$
coupling is suppressed, there will be a further increase in
BR$(\phi_W \tgg)$ which enhances the reach of the LHC in the $t\bar{t} \phi_W$
and $W \phi_W$ channels.

\item Large values for $A_b, A_t$ and $\mu$ with light bottom and 
top squarks may lead to a wide suppression of BR$(\phi_W\tgg)$, 
because the $\phi_W b\bar b$
coupling can be significantly enhanced.  This limits all of the 
$\phi_W\tgg$ channels at the LHC.
The upper bound on the Higgs boson mass is reduced in conjunction with
the increase in BR$(\phi_W\tbb)$, so that LEP and the Tevatron cover most
of the complementary regions of the $M_A-\tan\beta$ plane.  High luminosity
is only required at the Tevatron to cover the regions where 
neither CP-even Higgs boson has SM-like couplings to the 
gauge bosons, $\sin^2(\beta-\alpha) \simeq \cos^2(\beta-\alpha)$.

\item For MSSM
parameter choices 
where the Higgs
mixing would cause a suppression of both the $\phi_W b\bar b$ and
$\phi_W\tau^+\tau^-$ couplings at tree--level, large radiative corrections
from SUSY--breaking effects can modify the bottom and tau decay rates 
in different ways.  
As a result, one may observe 
$\phi_W(\to b\bar b)$ without $\phi_W(\to \tau^+\tau^-)$.
In the regions where there is a suppression of the $\phi_W\tau^+\tau^-$
coupling, there is not necessarily an enhancement of BR($\phi_W\tgg$),
because the $\phi_Wb\bar b$ coupling can be of the order of the
Standard Model value.
In the presence of large SUSY--breaking
effects, the suppression of the $\phi_W b\bar{b}$ coupling occurs when
$\tan\alpha\simeq \Delta(m_b)/\tan\beta$.  For this value of $\tan\alpha$,
the $\phi_W\tau^+\tau^-$ coupling will not vanish, in general, and 
$\phi_W\to\tau^+\tau^-$ may be the dominant decay.

\eit

While our analysis has emphasized the complementarity of LEP and the
Tevatron in the $\phi_W(\to b\bar b)$ channels to the LHC in the
$\phi_W(\tgg)$ channels, our results are more general.
If the experiments at the Tevatron do not receive
enough luminosity, then the 
$t\bar t\phi_W(\to b\bar b)$ channel, which to a good
approximation has the same parameter dependence as the
$W\phi_W(\to b\bar b)$ channel, at the LHC will be
complementary to the $\phi_W(\tgg)$ channels (the
$WW\to\phi_W(\to\tau^+\tau^-)$ channel may also be 
useful).  On the other hand, with enough luminosity,
the Tevatron may be able to observe the $\phi_W(\tgg)$
channel when the $\phi_Wb\bar b$ coupling is greatly
suppressed.  Regardless which of these scenarios is realized,
the next generation of measurements at LEP, the Tevatron
and the LHC can most likely reveal the nature of electroweak
symmetry breaking by observing or excluding a light Higgs boson
with Standard Model--like couplings to the $W$ and $Z$ bosons.

\medskip

{\bf{Acknowledgements}}
We thank D. Denegri and E. Richter--Was for providing 
information about various
channels.  We would also like to thank G. Azuelos, 
J. Gunion, H. Haber and J.~Womersley for interesting discussions. 
M.C. and C.W.
would like to thank the Aspen Center for Physics, where
part of this work has been done.
Work supported in part by the U.S. Department of Energy, High Energy
Physics Division, under Contract W-31-109-Eng-38.

\appendix
\section{Relations between the CP even Higgs masses}
\label{appendix_a}
To derive Eq.~(\ref{relation}),
we start with the Higgs squared-mass matrix elements 
parametrized in
the following form  (see Eq. (\ref{matrel})),
\begin{eqnarray}
{\cal M}^2_{11} & = & m_A^2 \sin^2\beta + \Delta_{11}
\nonumber\\
{\cal M}^2_{22} & = & m_A^2 \cos^2\beta + \Delta_{22}
\nonumber\\
{\cal M}^2_{12} & = &- m_A^2 \cos\beta\sin\beta + \Delta_{12},
\end{eqnarray}
where $\Delta_{ij}$ denotes terms independent of $m_A^2$.
For very large values of $m_A^2$, the heaviest CP-even Higgs
mass $m_H^2 \simeq m_A^2$ and the determinant of the Higgs
squared mass matrix will be equal to 
$m_A^2 \times m_h^2|_{m_A^2 \gg M_Z^2}$, where the last term
is the upper bound on the lightest CP-even Higgs mass.
This upper bound can be obtained by taking the terms
proportional to $m_A^2$ in the determinant of the Higgs
squared-mass matrix,
\begin{eqnarray}
\left.
m_h^2\right|_{m_A^2 \gg M_Z^2} & \simeq &
\Delta_{11} \cos^2\beta + \Delta_{22} \sin^2\beta +
2 \Delta_{12} \cos\beta\sin\beta 
\nonumber\\
& = & {\cal M}^2_{11} \cos^2\beta + {\cal M}^2_{22} \sin^2\beta
+ 2 {\cal M}^2_{12} \cos\beta \sin\beta 
\label{mijmh}
\end{eqnarray}

Since the mass matrix is diagonalized by a rotation with
mixing angle $\alpha$, we have
\begin{eqnarray}
{\cal M}_{11}^2 & = & m_h^2 \sin^2\alpha + m_H^2 \cos^2\alpha
\nonumber\\
{\cal M}_{22}^2 & = & m_h^2 \cos^2\alpha + m_H^2 \sin^2\alpha
\nonumber\\
{\cal M}_{12}^2 & = & \left(m_H^2-m_h^2\right) \sin\alpha \cos\alpha.
\label{mij}
\end{eqnarray} 

Substituting Eq. (\ref{mij}) into Eq. (\ref{mijmh}) yields
the desired relation, namely,
\begin{equation}
\left.
m_h^2\right|_{m_A^2 \gg M_Z^2} = m_h^2 \sin^2(\beta-\alpha) +
m_H^2 \cos^2(\beta - \alpha).
\label{hHboundh}
\end{equation}
Let us emphasize that, in the above, we have ignored the
small differences between
pole masses and running masses, while we have defined
all matrix elements at the scale $m_t$, ignoring the effects
of the decoupling of the heavy Higgs doublet. These effects,
however, are only relevant for $m_A \gg M_Z$, in
which case $\sin^2(\beta -\alpha) \rightarrow 1$ independently of the scale of 
definition. It is easy to prove that 
$\cos^2(\beta-\alpha) = {\cal O}(M_Z^4/m_A^4)$ for the
same conditions, and therefore the above equality,
Eq. (\ref{hHboundh}), is satisfied in a straightforward way.

%%%%%%%%%%%%%%%%%%%%%%%%%%%%%%%%%%%%%%%%%%%%%%
%
%  Begin Figures
%
%%%%%%%%%%%%%%%%%%%%%%%%%%%%%%%%%%%%%%%%%%%%%%%%
%%%%%%%%%%%%%%%%%%%%%%%%%%%%%%%%%%%%%%%%%%%%%%%%
\bce
\begin{figure}
\leavevmode
\bce
\epsfxsize=6.5in
\epsffile{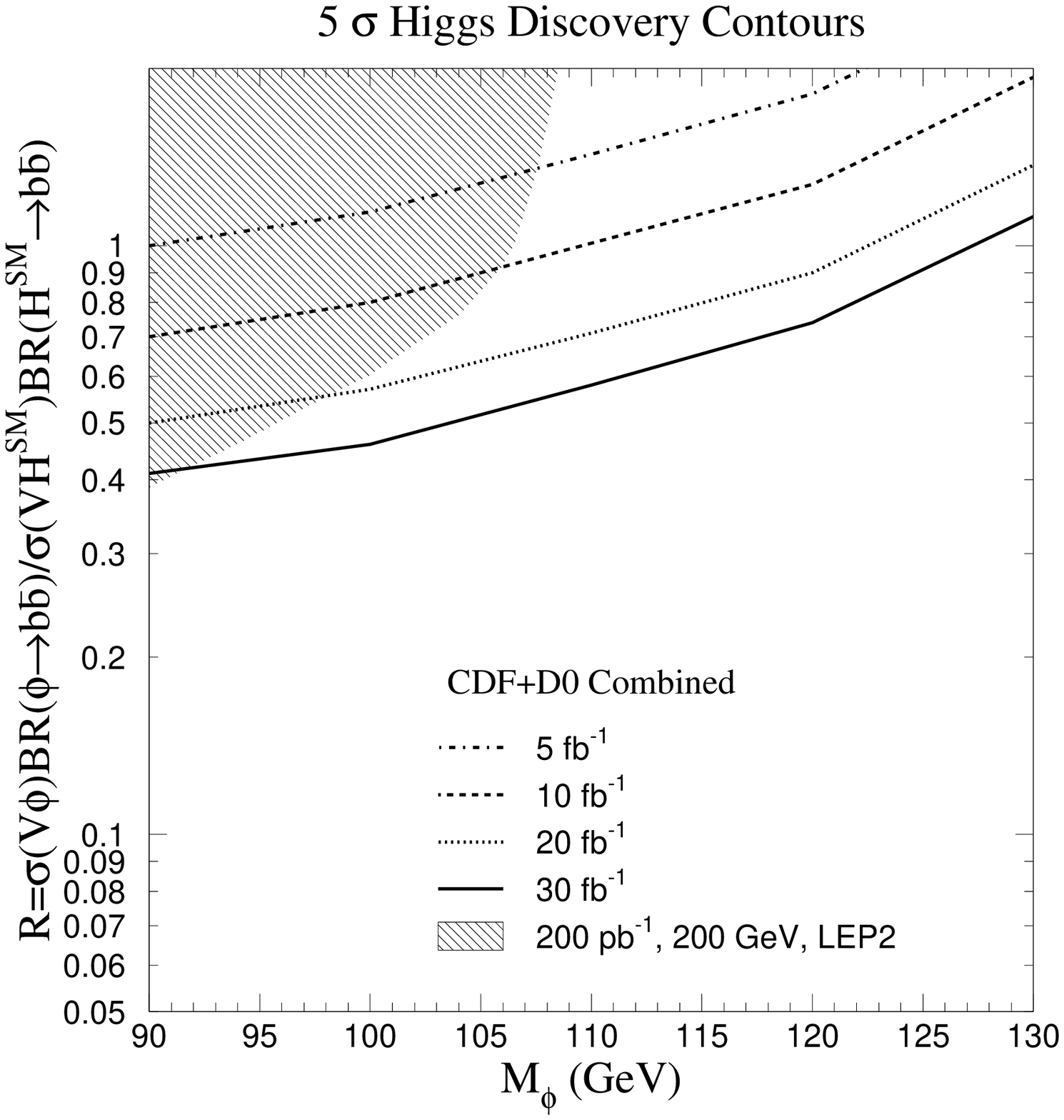}
\ece
\caption[]{Sensitivity of the Standard Model Higgs searches
at LEP and the Tevatron (for different total integrated luminosity).}
\label{rvalue}
\end{figure}
%%%%%%%%%%%%%%%%%%%%%%%%%%%%%%%%%%%%%%%%%%%%%%%%
\begin{figure}
\leavevmode
\bce
\epsfxsize=11cm
\epsffile{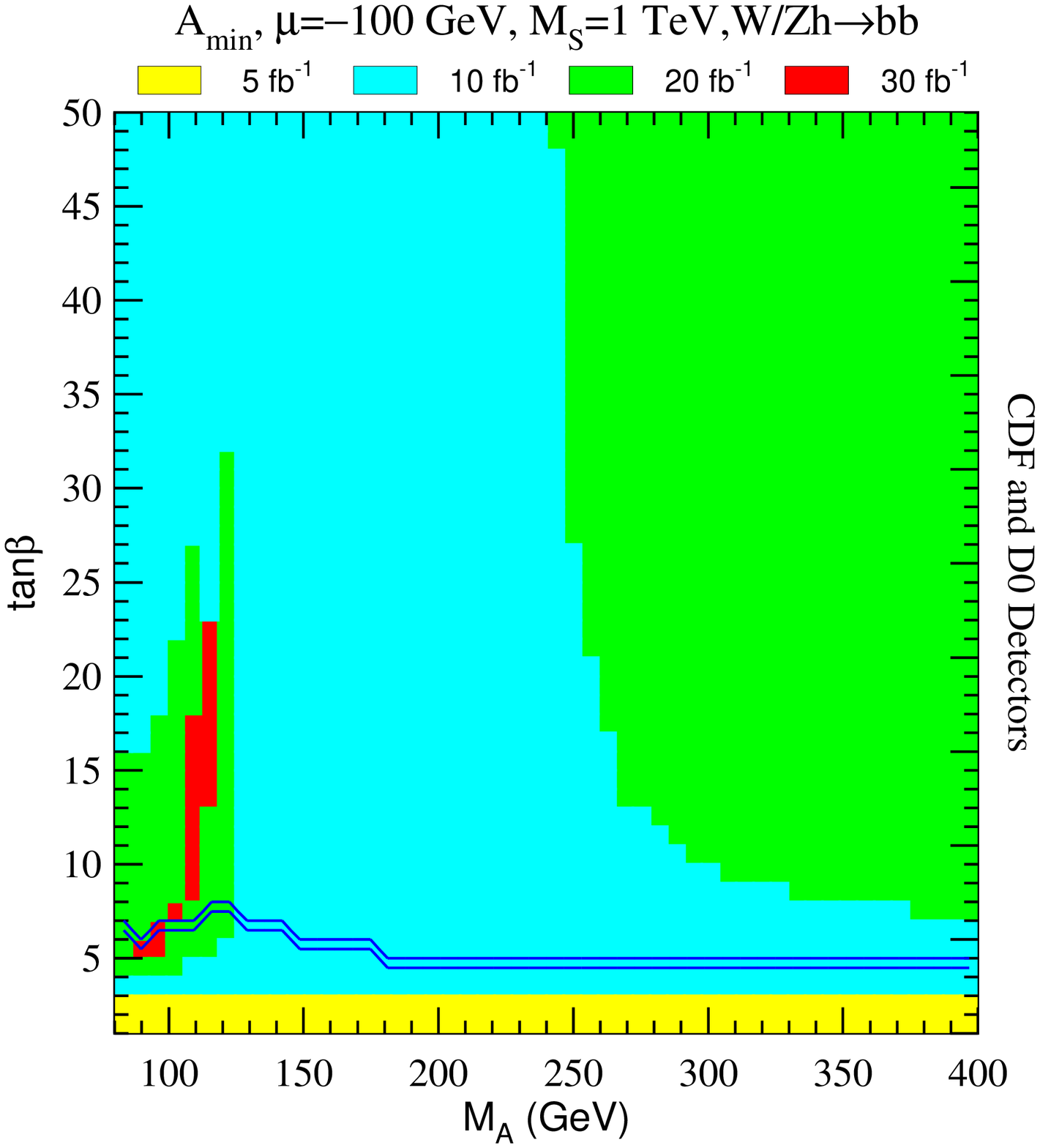}
\ece

\leavevmode
\bce
\epsfxsize=11cm
\epsffile{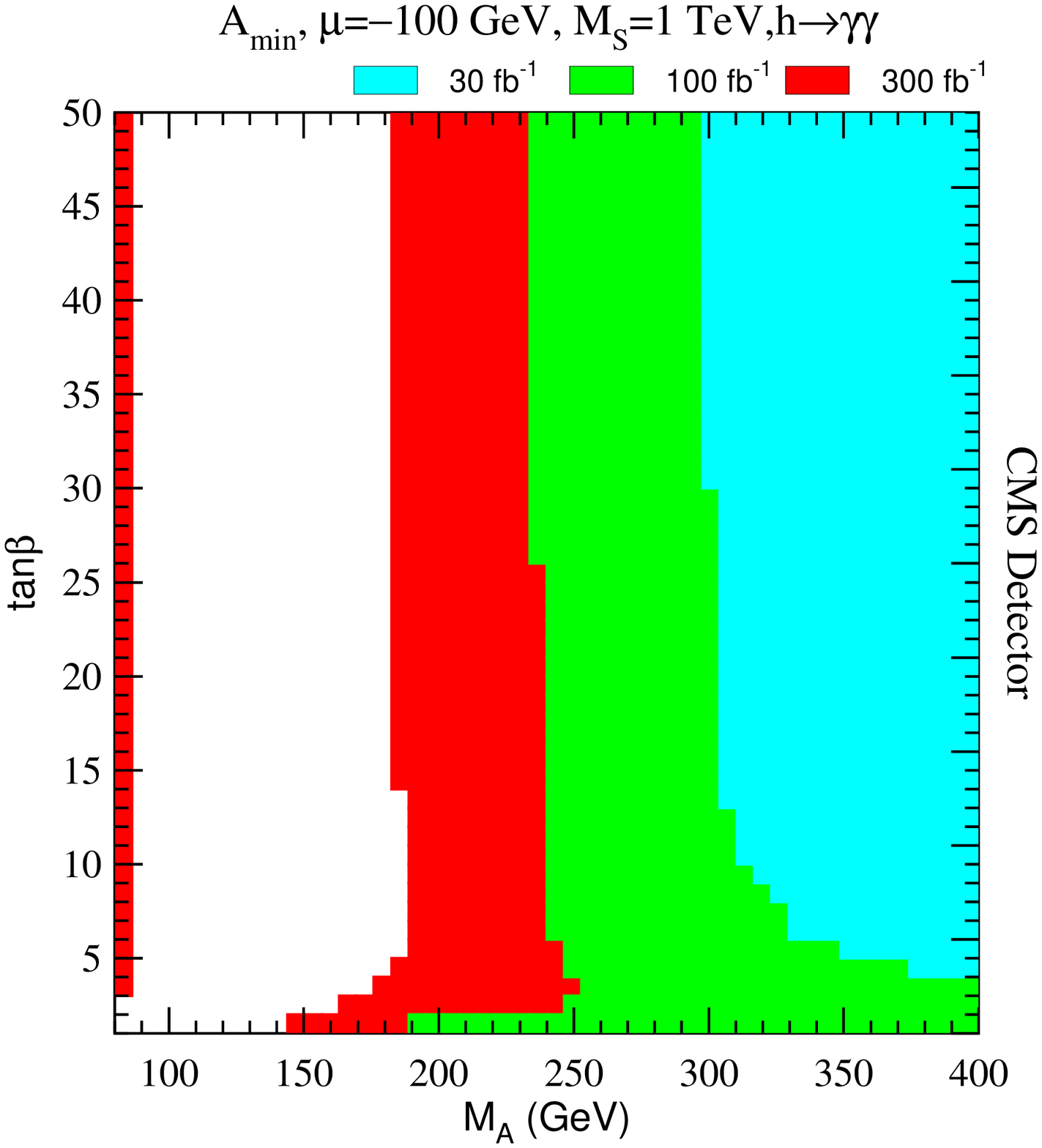}
\ece
\caption[]{Discover reach of the LEP,
Tevatron and LHC experiments in
the minimal mixing model, as defined in the text.}
\label{minimal}
\end{figure}
%%%%%%%%%%%%%%%%%%%
\begin{figure}
\leavevmode
\bce
\epsfxsize=11cm
\epsffile{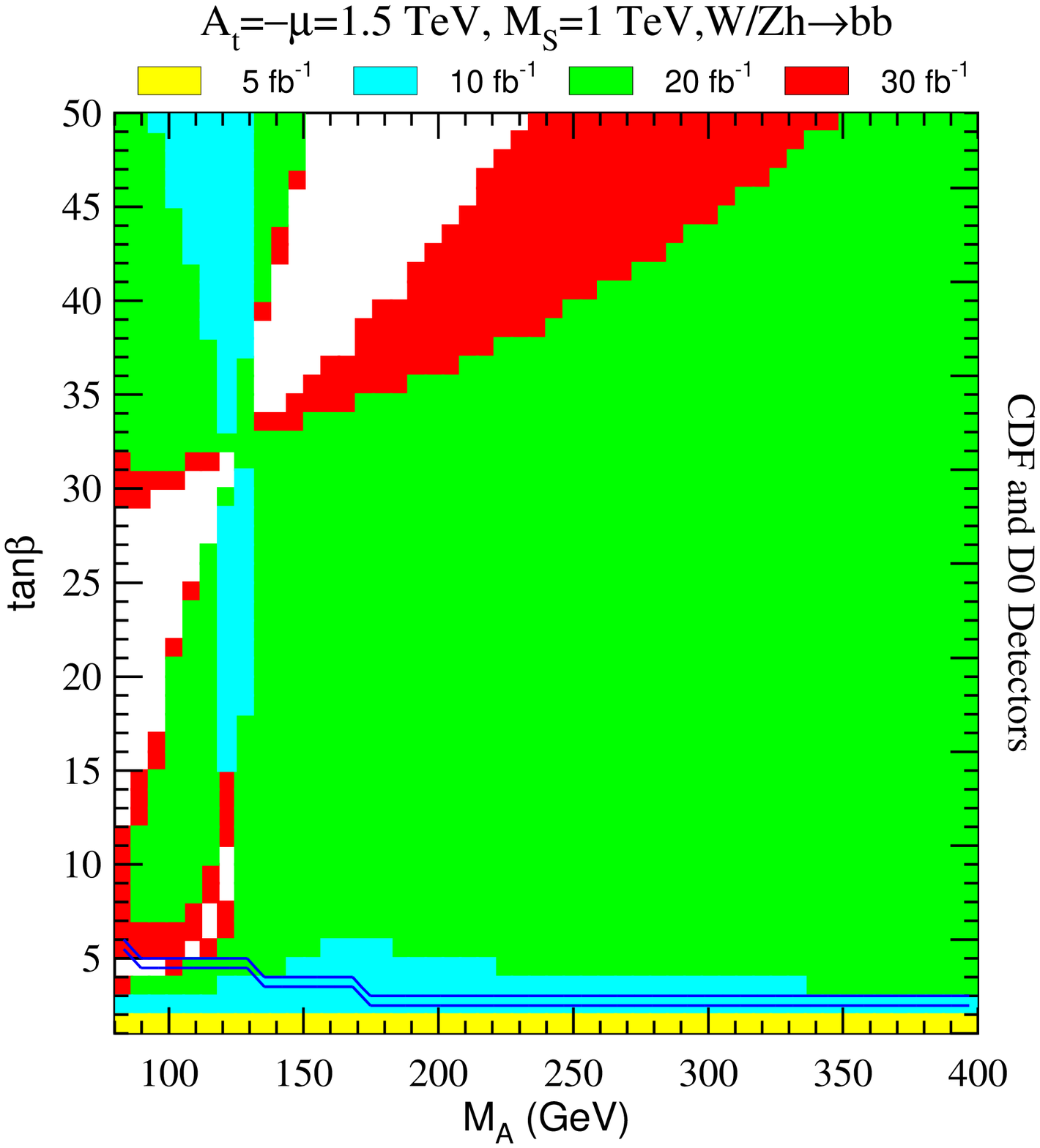}
\ece

\leavevmode
\bce
\epsfxsize=11cm
\epsffile{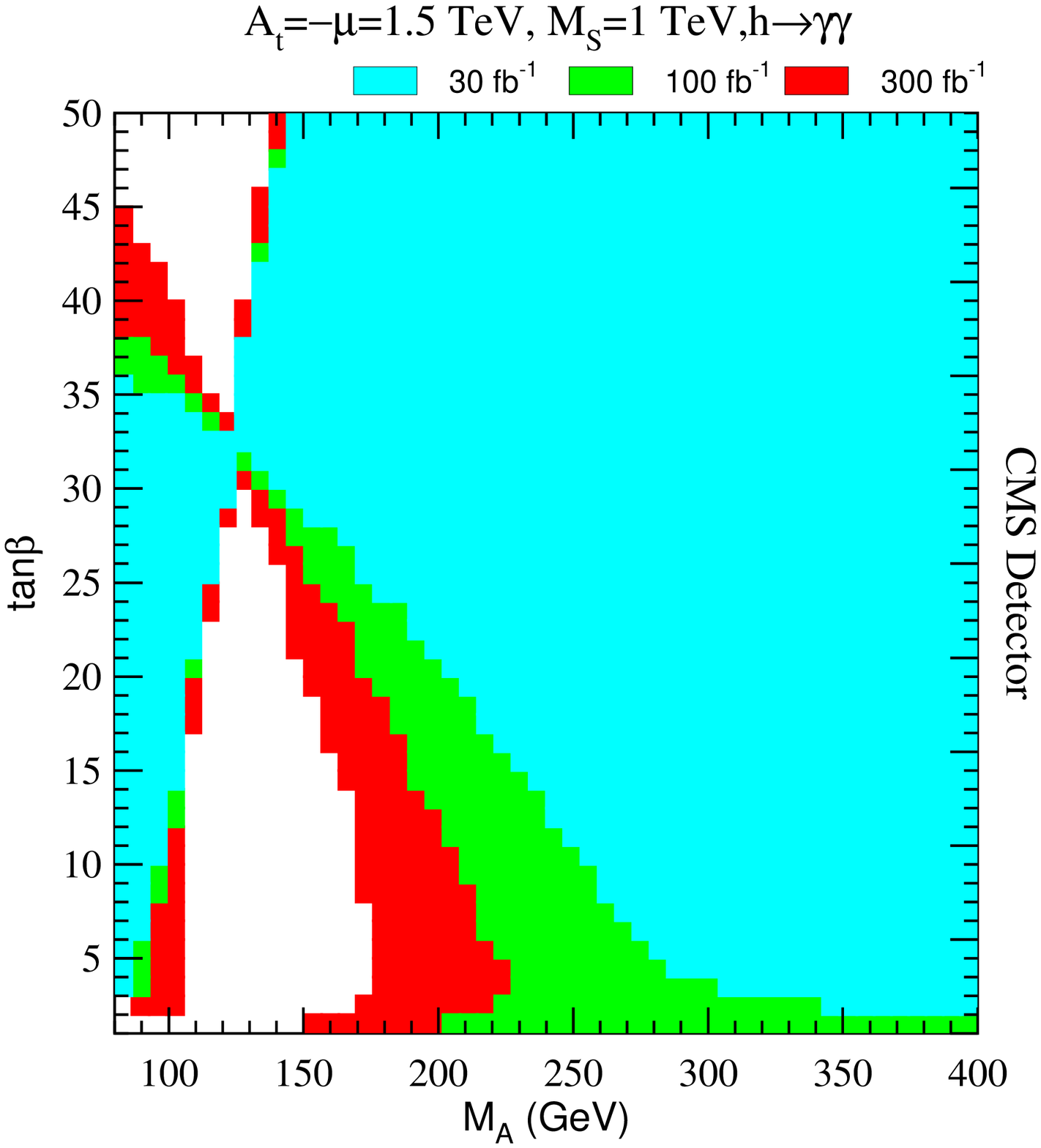}
\ece
\caption[]{Same as Fig.~\ref{minimal} but for
$A_t=-\mu=1.5$ TeV, $M_S=1$ TeV}
\label{mix_large_ms}
\end{figure}
%%%%%%%%%%%%%%%%%%%
%%%%%%%%%%%%%%%%%%%
\begin{figure}
\leavevmode
\bce
\epsfxsize=11cm
\epsffile{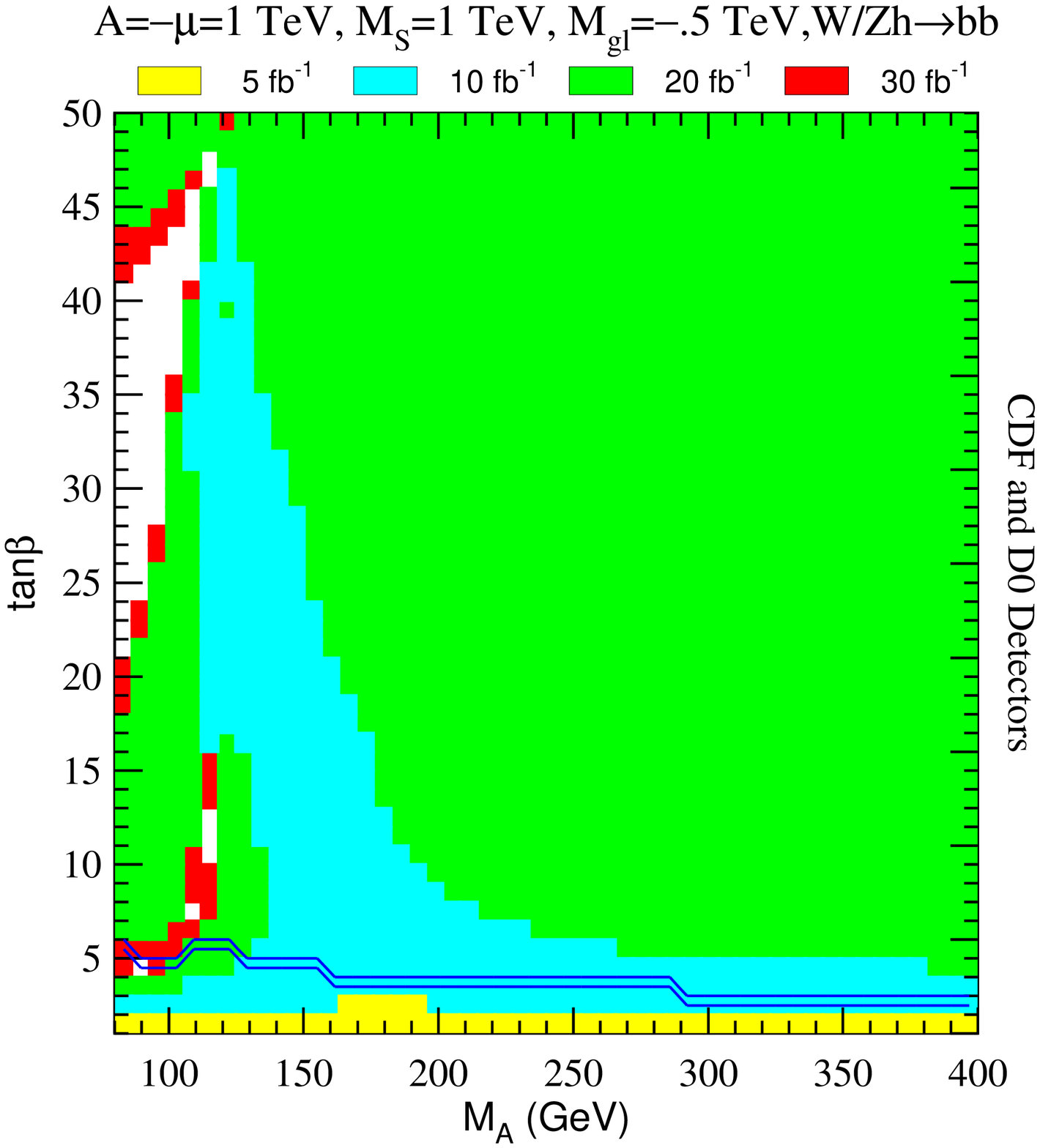}
\ece

\leavevmode
\bce
\epsfxsize=11cm
\epsffile{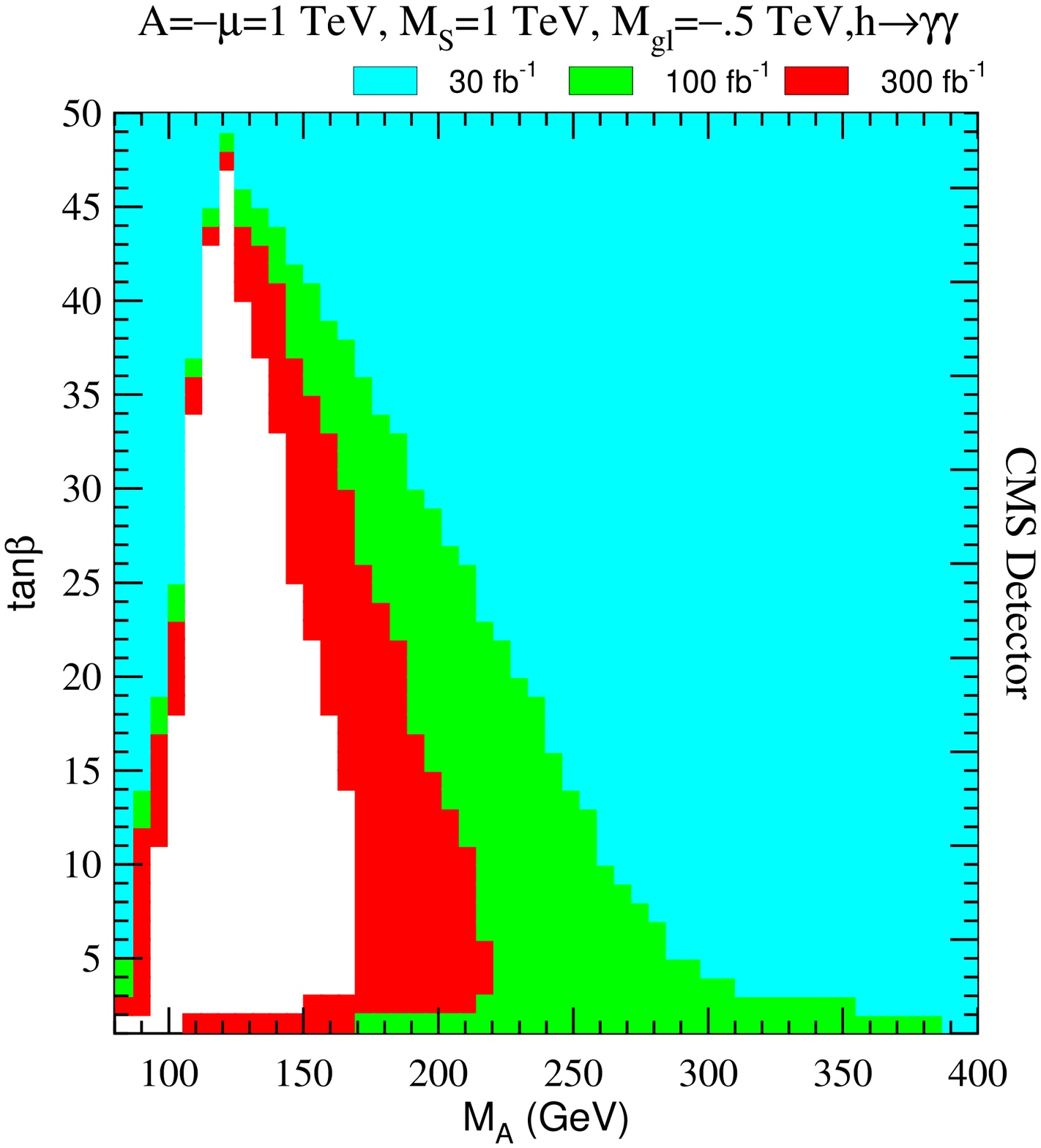}
\ece
\caption[]{Same as Fig.~\ref{minimal}, but for $A_t = -\mu = 1$ TeV,
and including the effects of the bottom mass corrections, 
$\Delta(m_b)$, calculated using
$M_{\tilde g}=-.5$ TeV.}
\label{deltamb1_neg_mg}
\end{figure}
%%%%%%%%%%%%%%%%%%%
\begin{figure}
\leavevmode
\bce
\epsfxsize=11cm
\epsffile{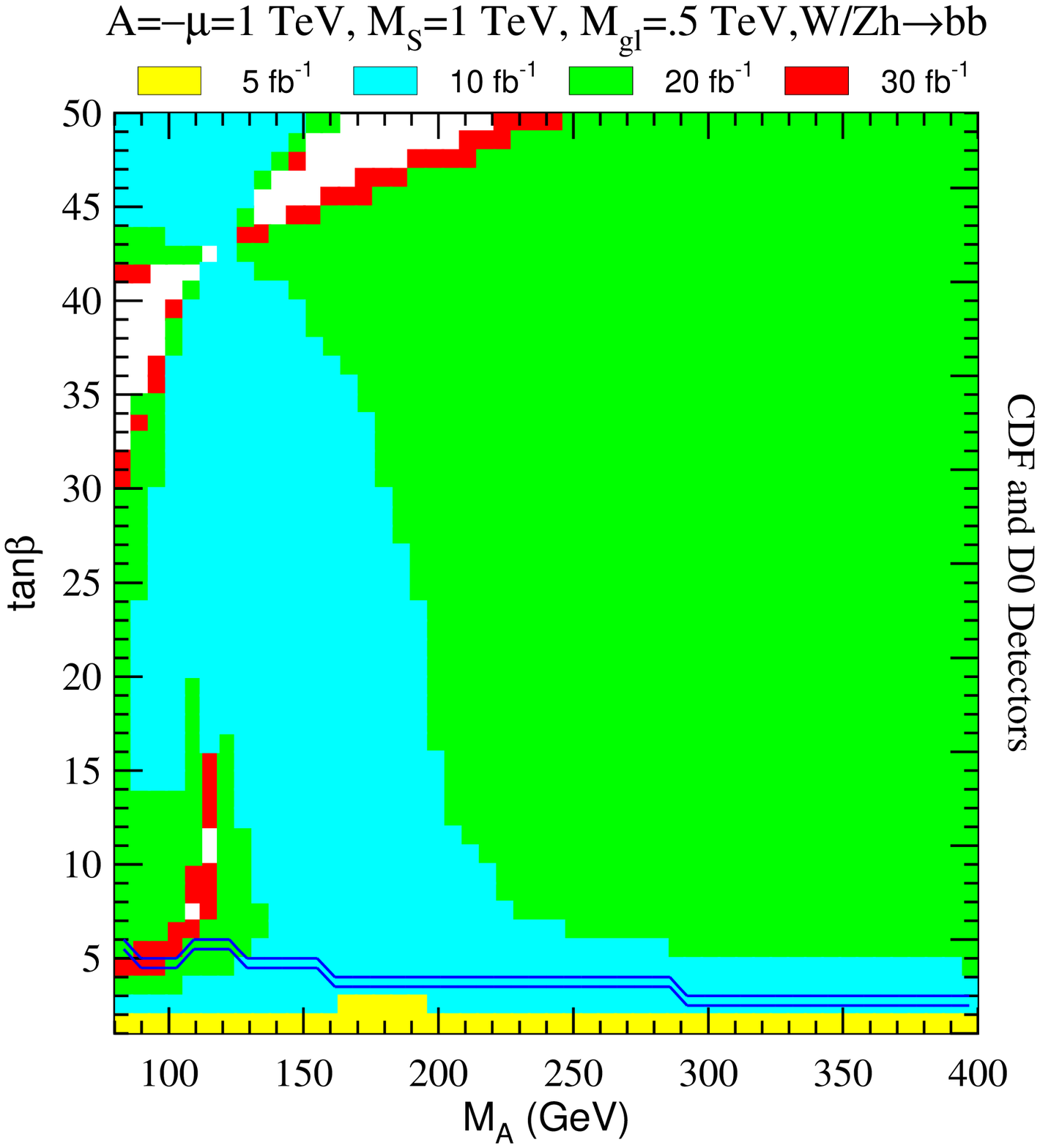}
\ece

\leavevmode
\bce
\epsfxsize=11cm
\epsffile{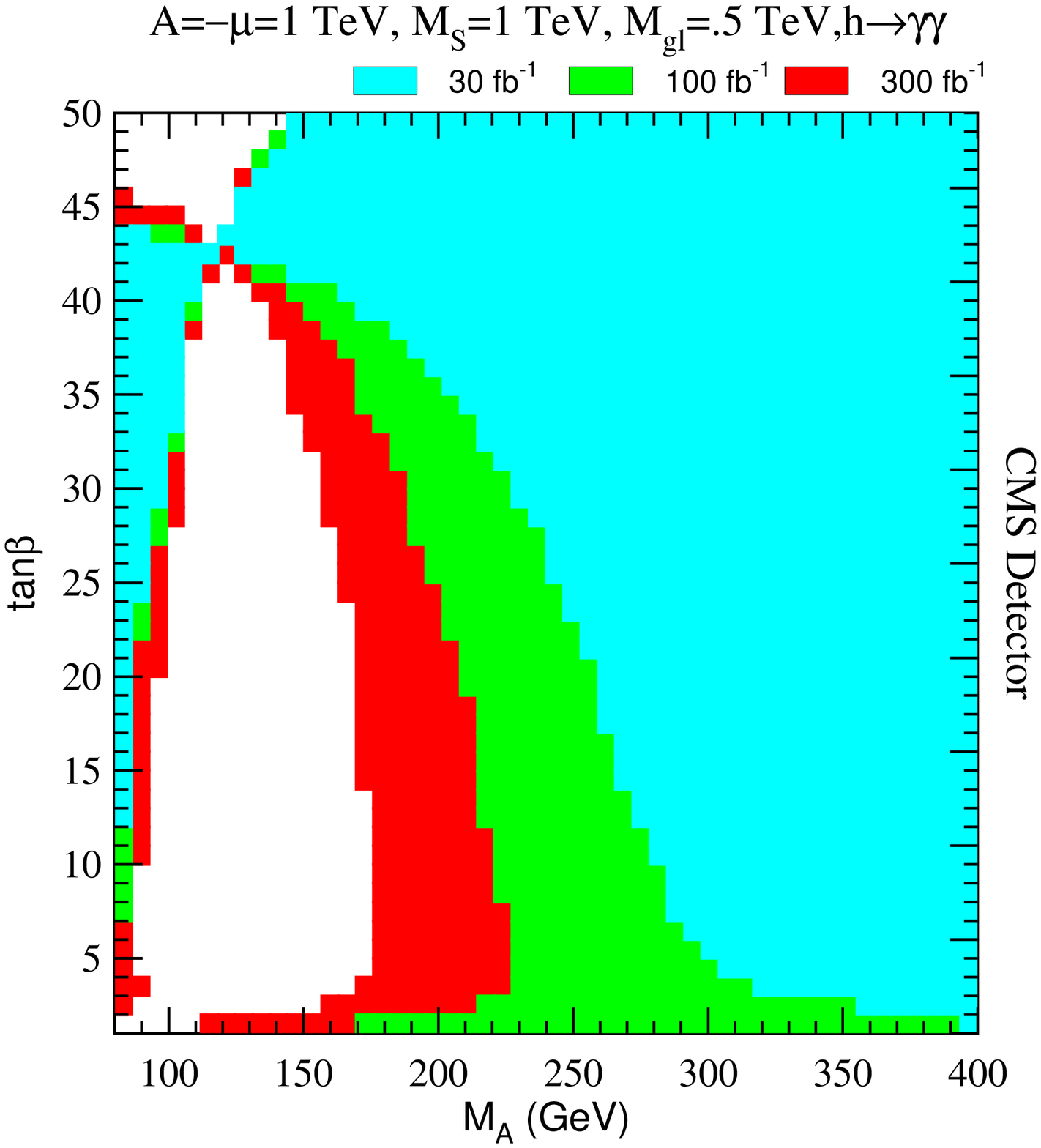}
\ece
\caption[]{Same as Fig.~\ref{deltamb1_neg_mg}, 
but for 
$M_{\tilde g}=.5$ TeV.}
\label{deltamb1_pos_mg}
\end{figure}
%%%%%%%%%%%%%%%%%%%%%%%%%%%%%%%%%%%%%
\begin{figure}
\leavevmode
\bce
\epsfxsize=11cm
\epsffile{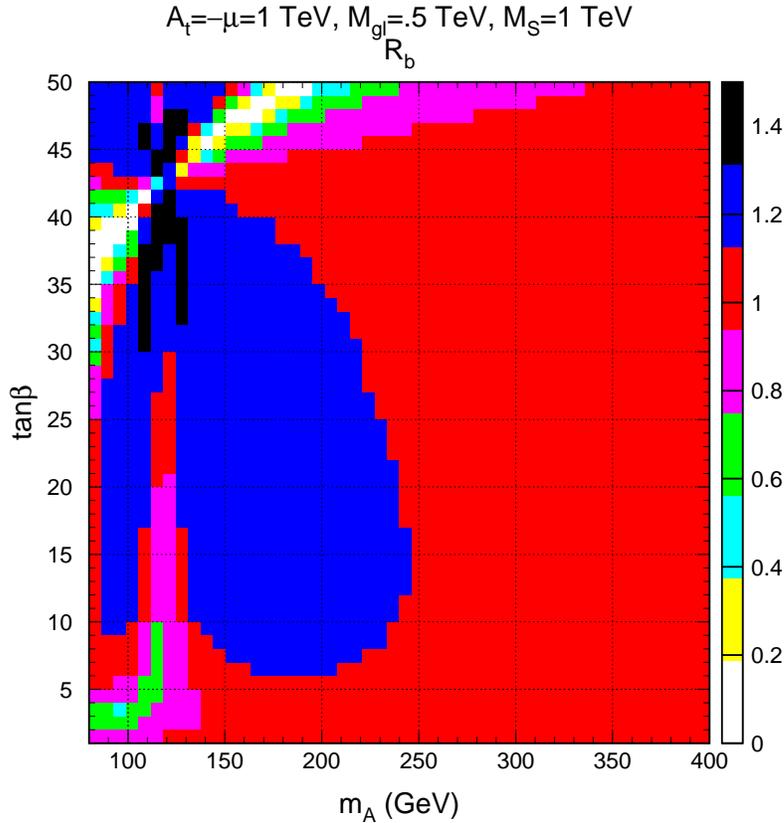}
\ece

\leavevmode
\bce
\epsfxsize=11cm
\epsffile{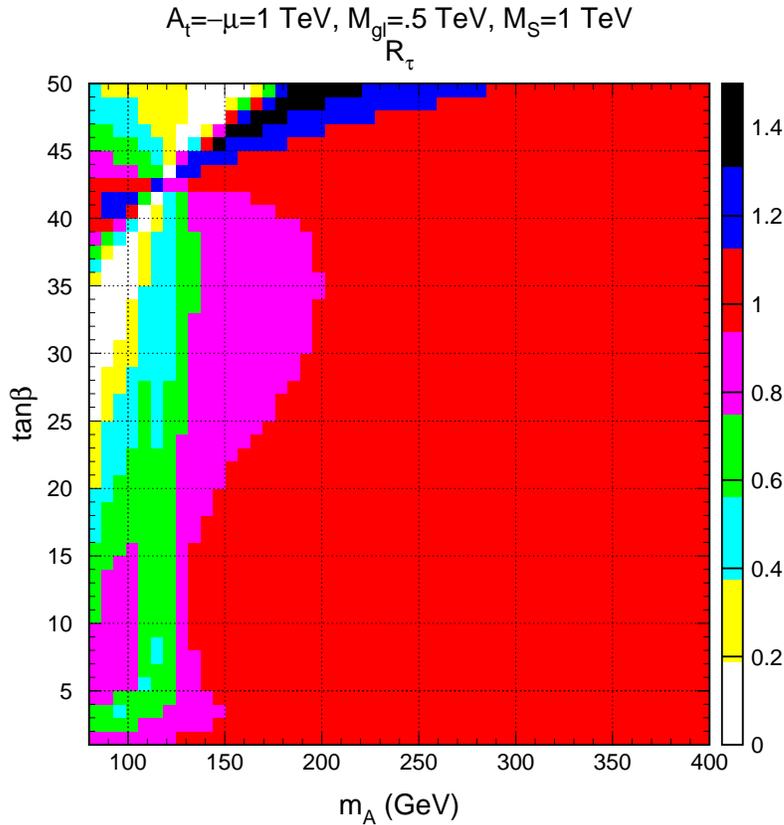}
\ece
\caption[]{Comparison of the sensitivity of the 
$\phi_W \rightarrow b\bar b$ and $
\phi_W \rightarrow \tau^+\tau^-$ channels,
$R_b$ and $R_\tau$, respectively, 
when the Higgs $\phi_W$ is produced from a $VV\phi_W$ vertex, $V=W$ or $Z$.}
\label{tau_vs_bbbar}
\end{figure}
%%%%%%%%%%%%%%%%%%%
\begin{figure}
\leavevmode
\bce
\epsfxsize=11cm
\epsffile{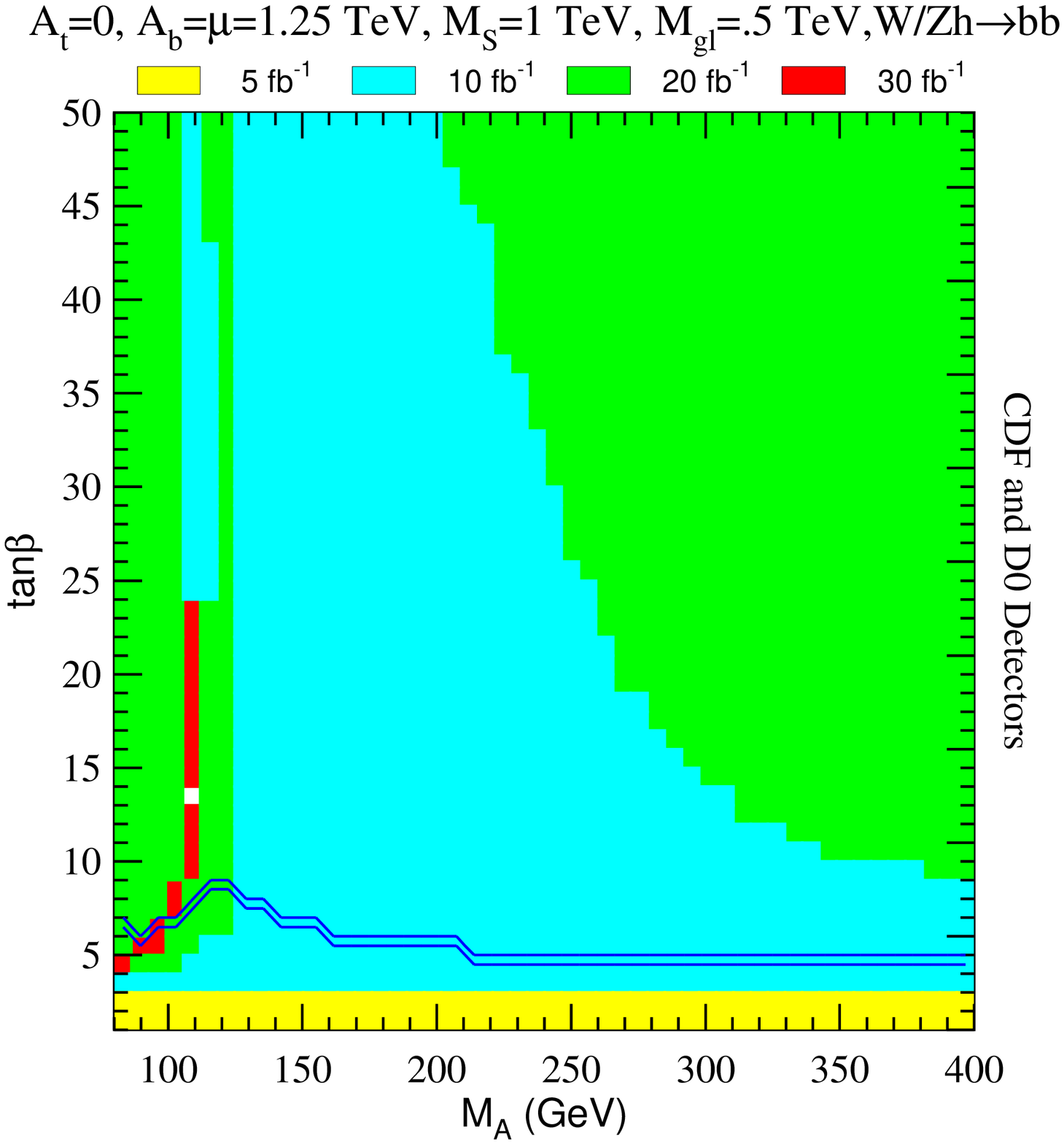}
\ece

\leavevmode
\bce
\epsfxsize=11cm
\epsffile{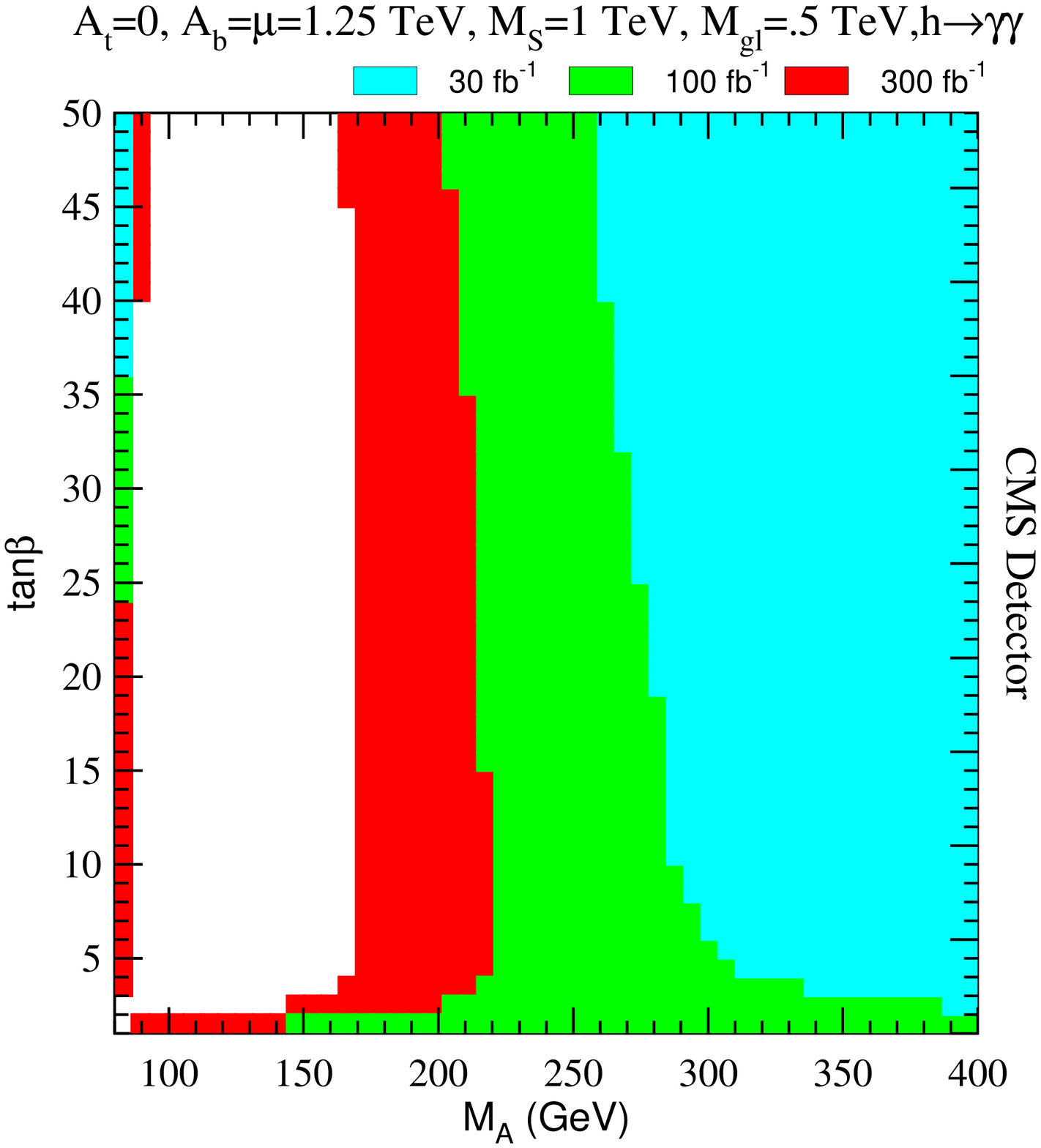}
\ece
\caption[]{Same as Fig.~\ref{minimal}, but considering a possible
cancellation of BR$(\phi_W \to b \bar b)$
for large $A_b=\mu$ and $\Delta(m_b)>0$.}
\label{ab_eq_mu_pos_mg}
\end{figure}
%%%%%%%%%%%%%%%%%%%
\begin{figure}
\leavevmode
\bce
\epsfxsize=11cm
\epsffile{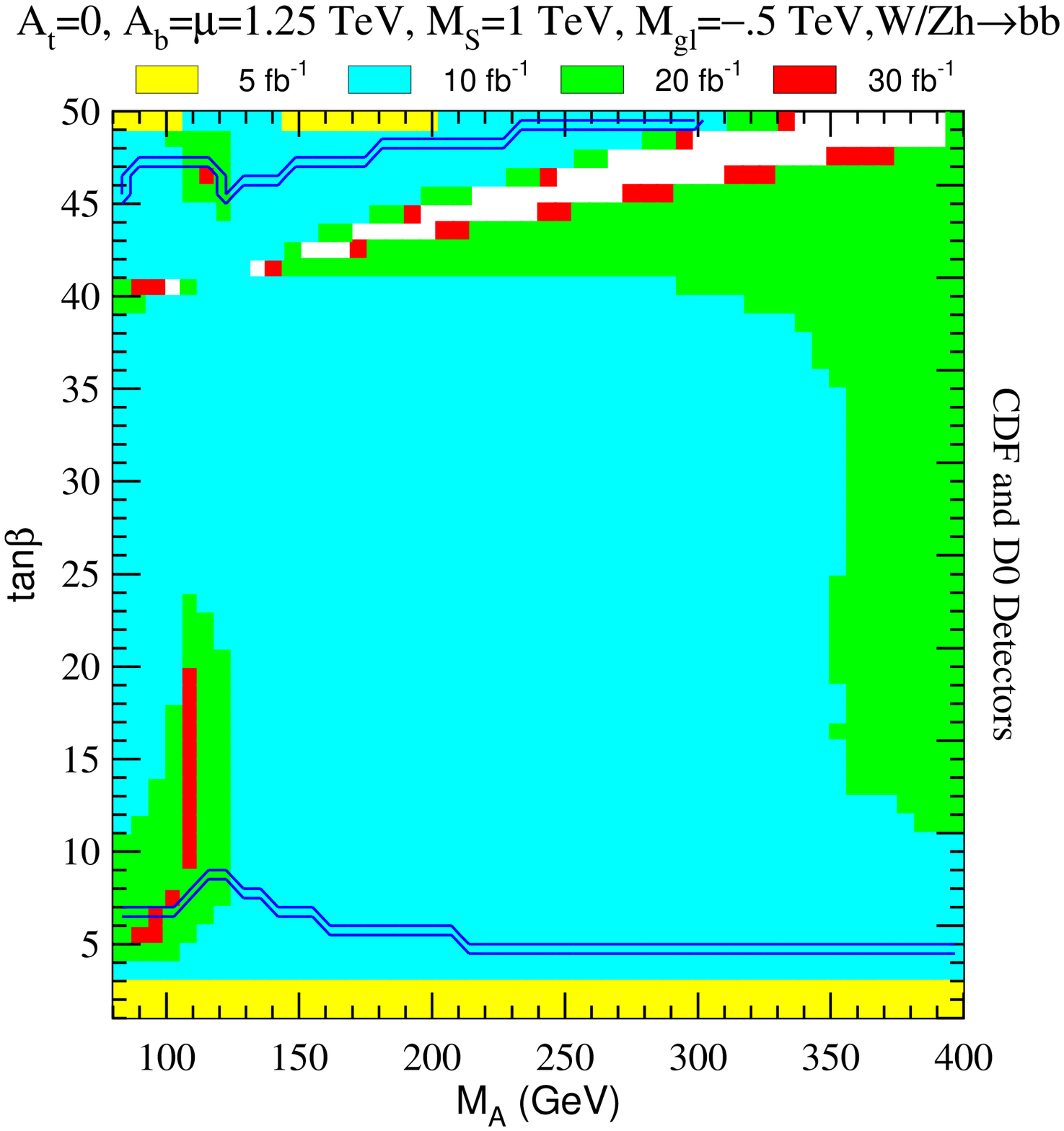}
\ece

\leavevmode
\bce
\epsfxsize=11cm
\epsffile{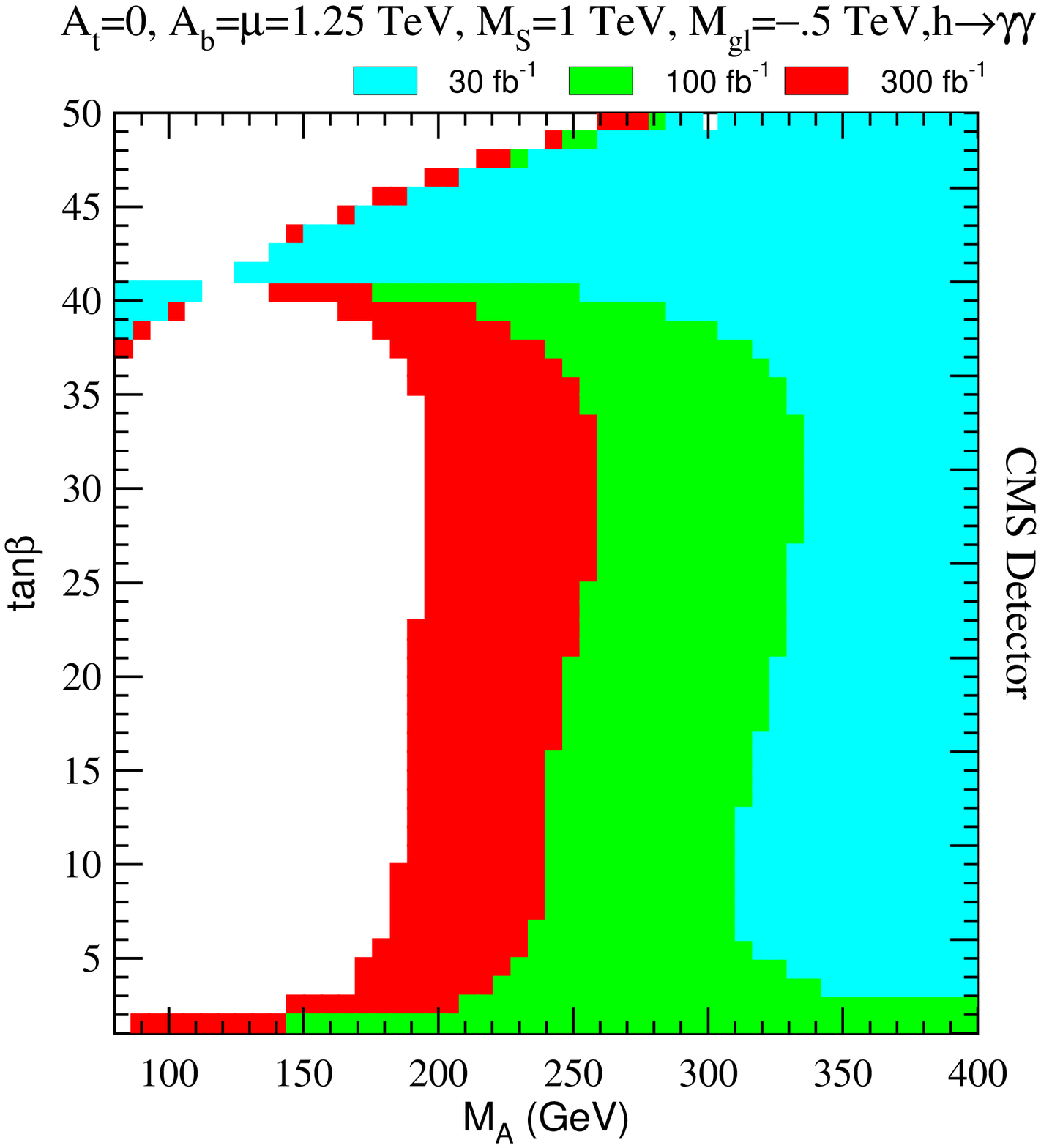}
\ece
\caption[]{Same as Fig~\ref{ab_eq_mu_pos_mg} but for
$\Delta(m_b)<0$.}
\label{ab_eq_mu_neg_mg}
\end{figure}
%%%%%%%%%%%%%%%%%%%
\begin{figure}
\leavevmode
\bce
\epsfxsize=11cm
\epsffile{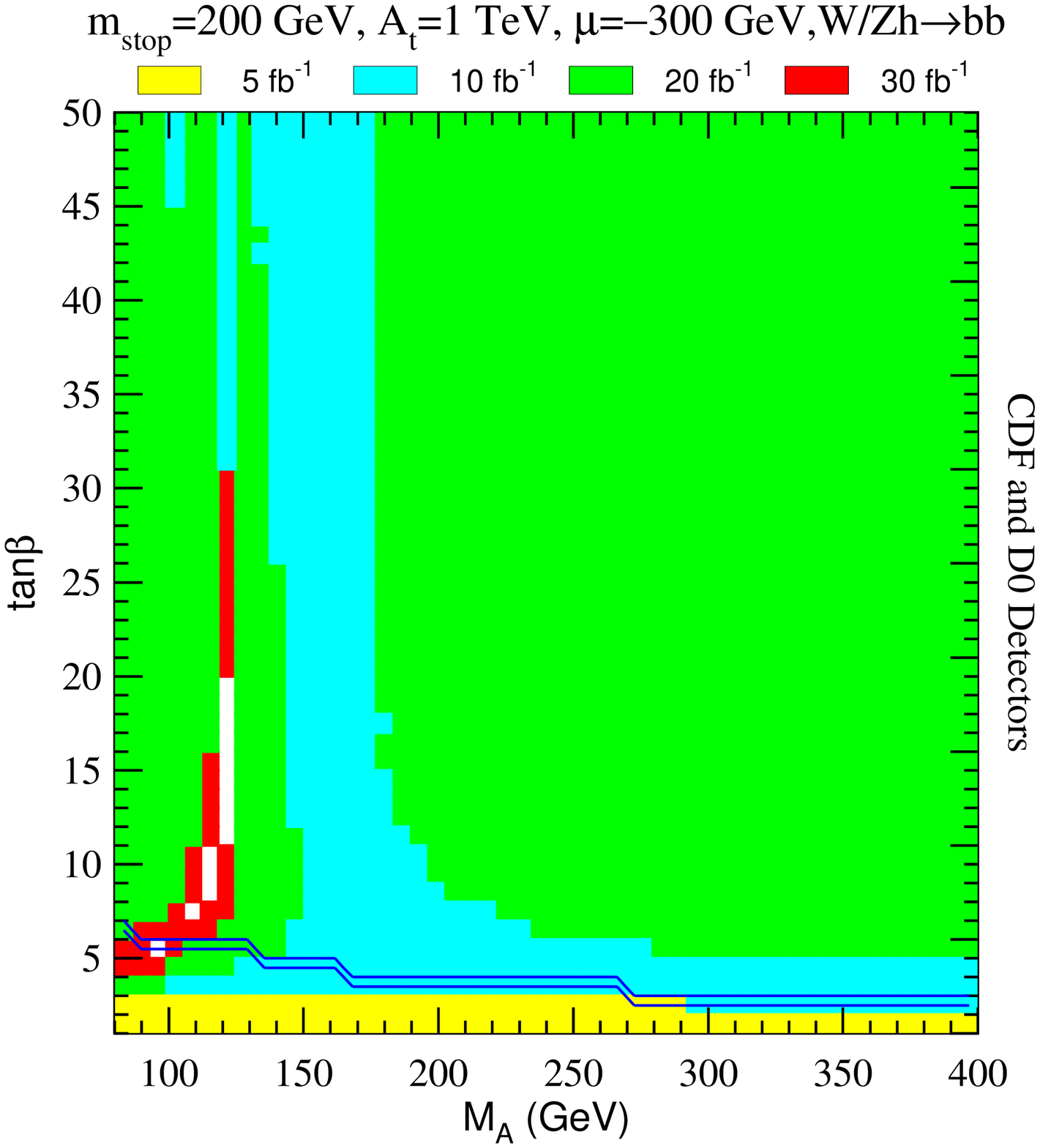}
\ece

\leavevmode
\bce
\epsfxsize=11cm
\epsffile{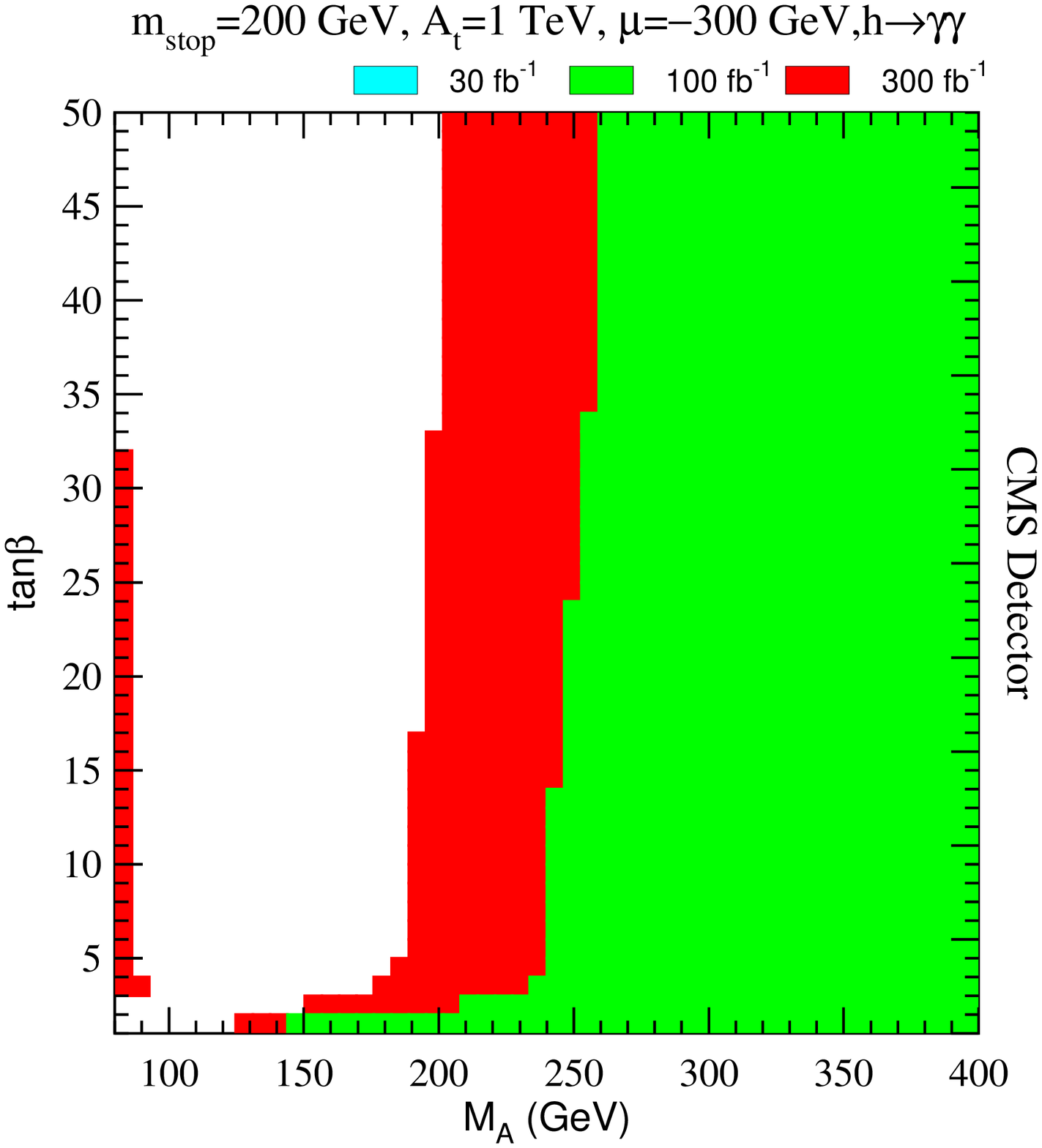}
\ece
\caption[]{Same as Fig~\ref{minimal}, but for a region of
parameters such that $\Gamma(\phi_W \rightarrow g g)$ is 
suppressed with respect to the SM value,
$M_{\tilde{t}_1}=200$ GeV, $A_t=1$ TeV, $\mu=-.3$ TeV}
\label{light_stop}
\end{figure}
%%%%%%%%%%%%%%%%%%%
\begin{figure}
\leavevmode
\bce
\epsfxsize=11cm
\epsffile{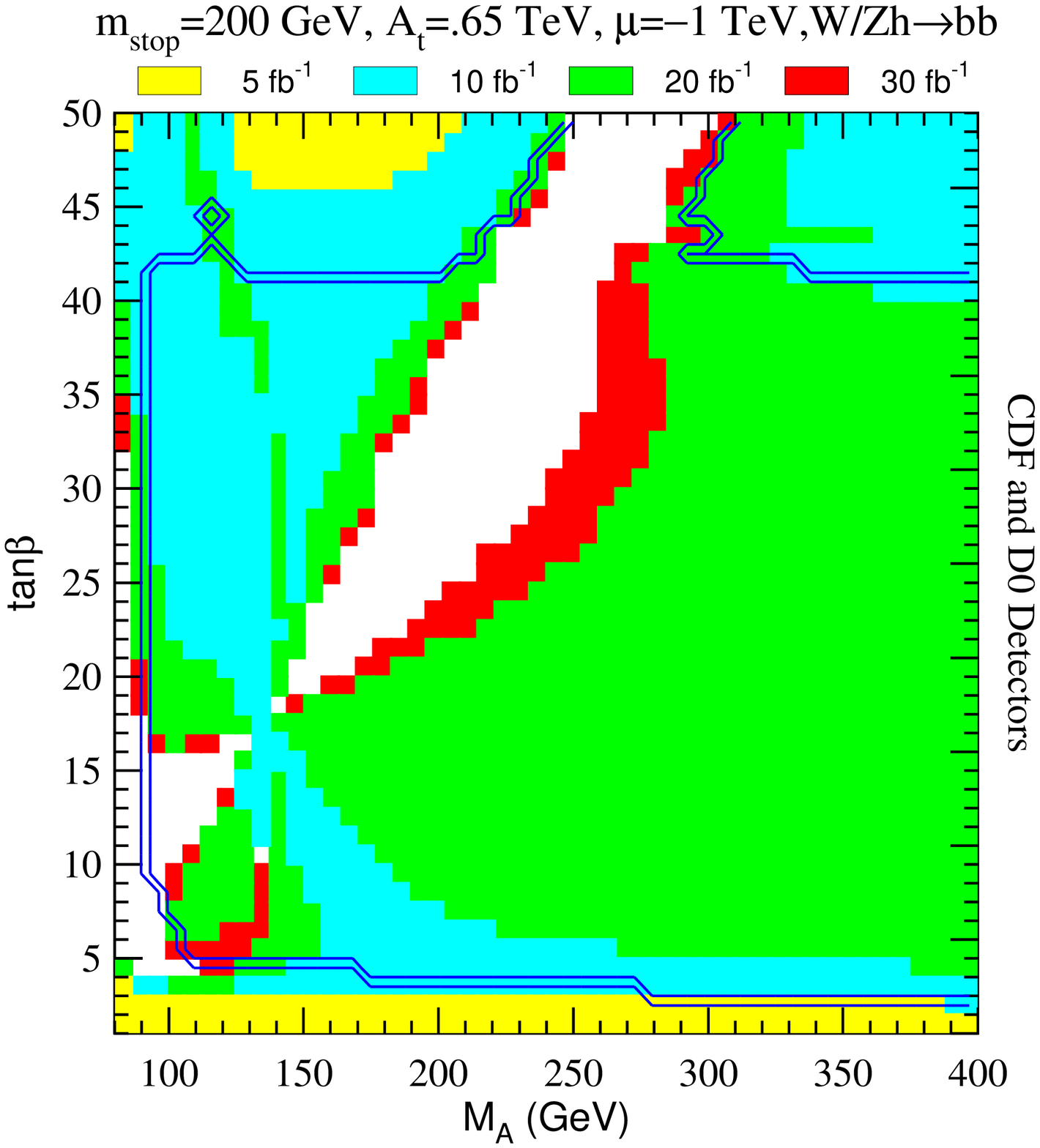}
\ece

\leavevmode
\bce
\epsfxsize=11cm
\epsffile{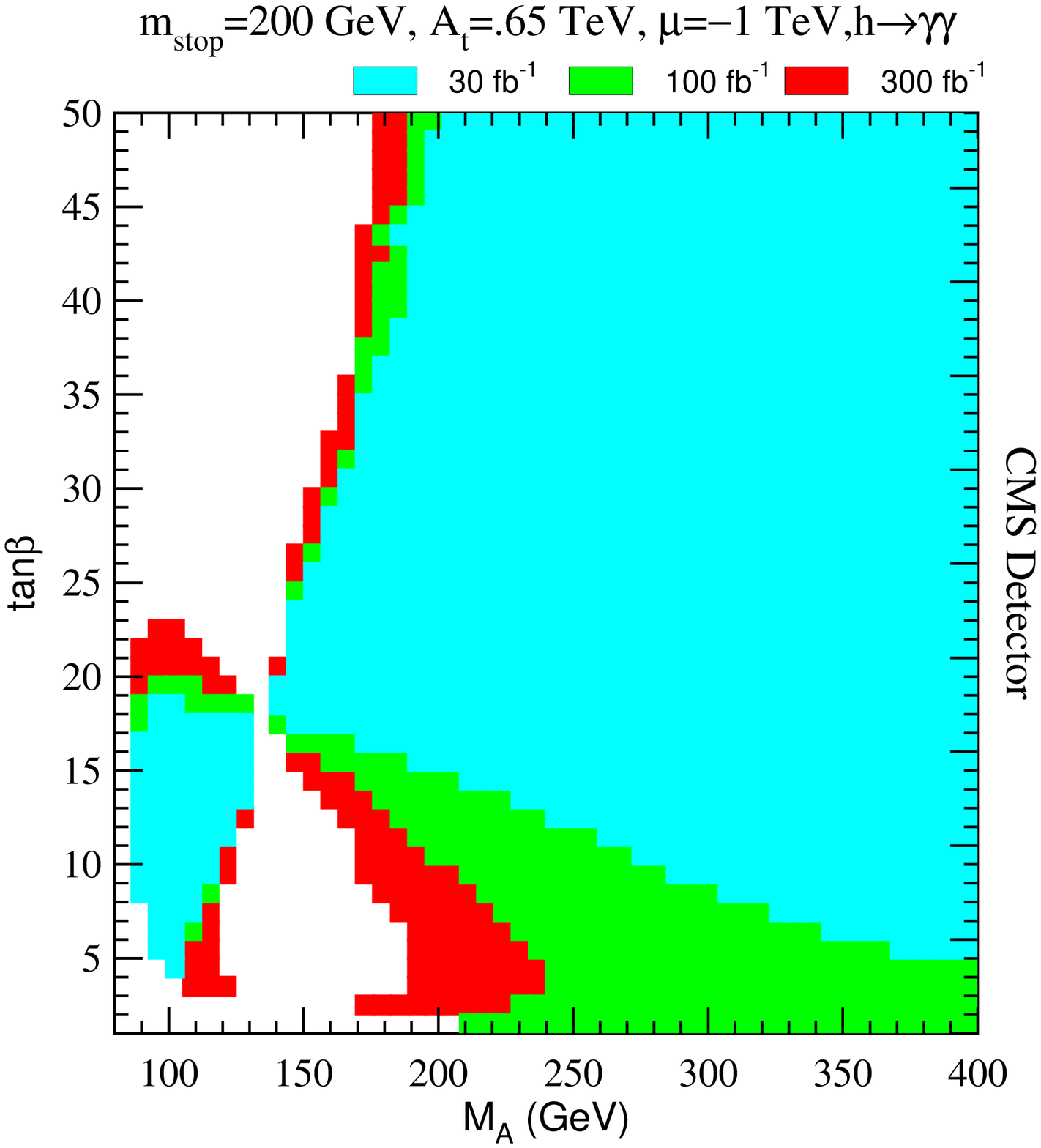}
\ece
\caption[]{Same as Fig.~\ref{light_stop}, but considering stop
mixing mass parameters
which induce a suppression of the $\phi_W b \bar{b}$
coupling,
$M_{\tilde{t}_1}=200$ GeV, $A_t=.65$ TeV, $\mu=-1$ TeV}
\label{mix_small_ms}
\end{figure}
%%%%%%%%%%%%%%%%%%%
\begin{figure}
\leavevmode
\bce
\epsfxsize=11cm
\epsffile{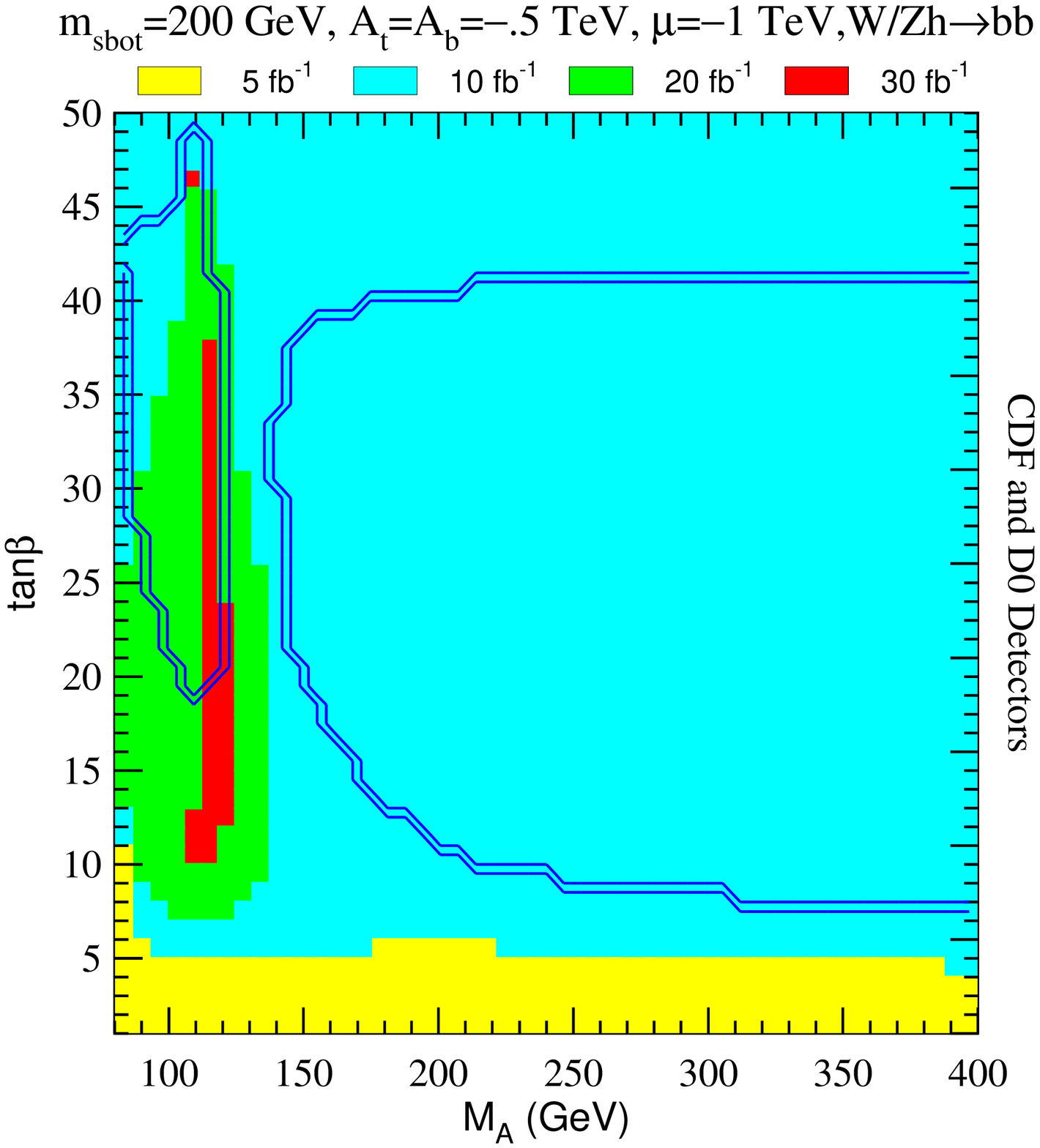}
\ece

\leavevmode
\bce
\epsfxsize=11cm
\epsffile{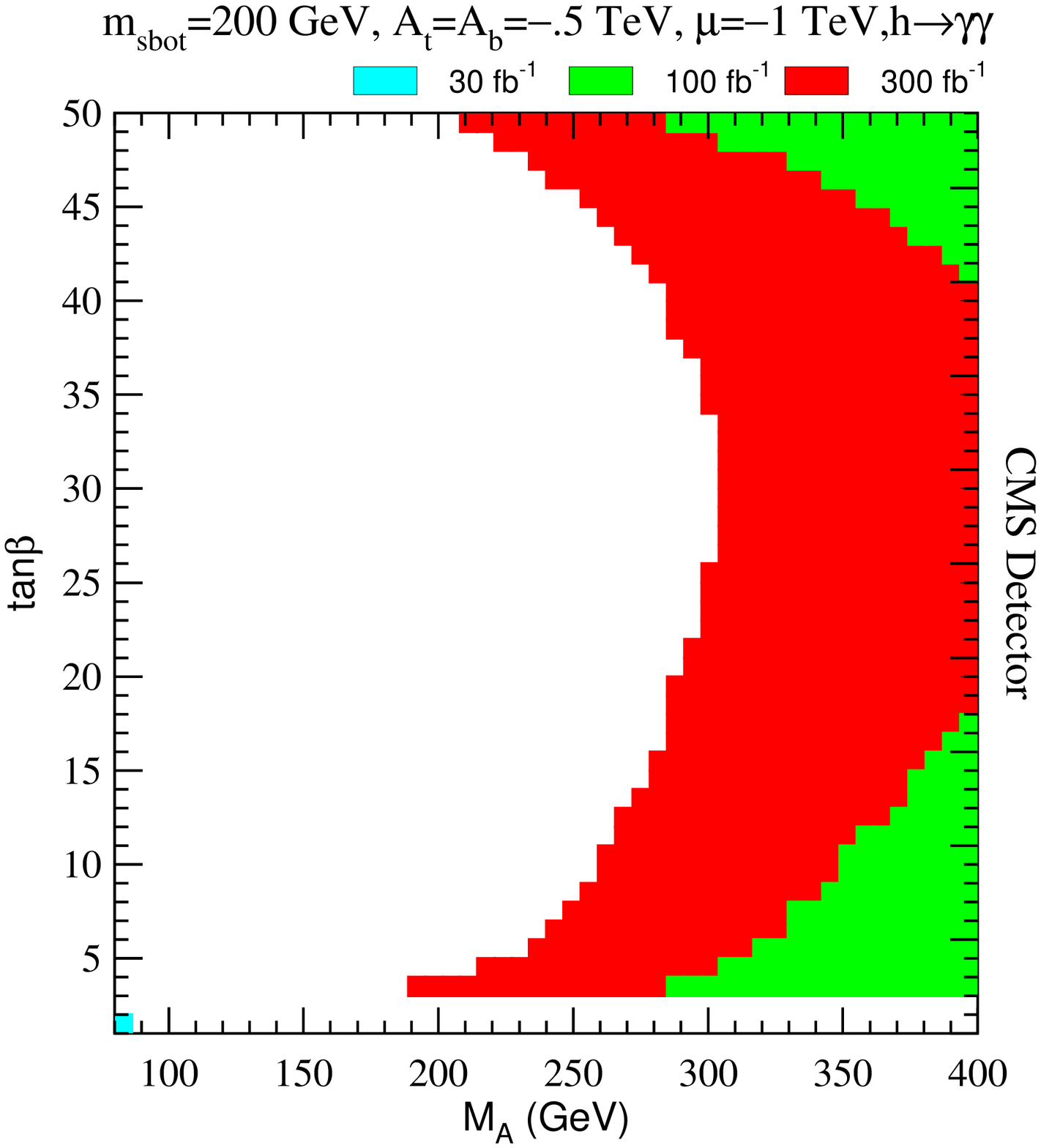}
\ece
\caption[]{Same as Fig.~\ref{minimal}, but for light top
and bottom squarks, and large mixing mass parameters,
$M_{\tilde{b}_1}=200$ GeV, $A_t=A_b=-.5$ TeV, $\mu=-1$ TeV}
\label{light_sbot}
\end{figure}
\ece
%%%%%%%%%%%%%%%%%%%
\end{document}
%
% ****** End of file apssamp.tex ******